\documentclass{interact}

\usepackage{epstopdf}
\usepackage[caption=false]{subfig}

\usepackage{array}
\usepackage{wrapfig}
\usepackage[T1]{fontenc}
\usepackage[utf8]{inputenc}
\DeclareUnicodeCharacter{2243}{\ensuremath{\simeq}}
\DeclareUnicodeCharacter{224D}{\ensuremath{\asymp}}
\DeclareUnicodeCharacter{2500}{--}
\DeclareUnicodeCharacter{0306}{}
\DeclareUnicodeCharacter{03B1}{\ensuremath{\alpha}}
\DeclareUnicodeCharacter{237A}{\ensuremath{\alpha}}
\DeclareUnicodeCharacter{03BC}{\ensuremath{\mu}}
\DeclareUnicodeCharacter{2609}{\ensuremath{\odot}}

\newcolumntype{C}[1]{>{\centering\arraybackslash}p{#1}}

\usepackage[numbers,sort&compress]{natbib}
\bibpunct[, ]{[}{]}{,}{n}{,}{,}
\makeatletter
\def\NAT@def@citea{\def\@citea{\NAT@separator}}
\makeatother

\usepackage[hypertexnames=false]{hyperref}
\usepackage[dvipsnames]{xcolor} 
\hypersetup{
    colorlinks=true,
    linkcolor=NavyBlue,    
    citecolor=NavyBlue,    
    urlcolor=NavyBlue,
    filecolor=NavyBlue
}

\theoremstyle{plain}

\theoremstyle{definition}

\theoremstyle{remark}

\newcommand{\arcsec}{^{\prime\prime}}

\begin{document}

\articletype{Introductory review article}

\title{{\itshape How {\it JWST} is Revolutionizing Our View of Early
    Galaxy Formation}}

\author{
  \name{Rychard Bouwens\textsuperscript{*}\thanks{Email: bouwens@strw.leidenuniv.nl} }
  \vspace{-0.2cm}
\affil{\textsuperscript{*}Leiden Observatory, University of Leiden, 2333CA Leiden, The Netherlands}
}

\maketitle

\begin{abstract}
 The {\it James Webb Space Telescope} ({\it JWST}) has been
 revolutionizing our understanding of early galaxy formation over its
 first four years of science operations. With unprecedented
 sensitivity, spatial resolution, and wavelength coverage from the
 near- to mid-infrared, {\it JWST} has enabled astronomers to identify
 hundreds of star-forming galaxies at redshifts z$\geq$8 (within
 $\sim$650 million years of the Big Bang) through strong spectral
 breaks and characterize these galaxies through sensitive
 spectroscopy. These capabilities have provided new insights into the
 stellar populations, neutral gas, nebular continuum emission,
 chemical enrichment, dust content, and ionization conditions of
 galaxies during cosmic dawn, revealing more numerous and brighter
 systems than anticipated from pre-{\it JWST} models. We discuss the
 implications of these discoveries for reionization, the buildup of
 stellar mass and metals, and the evolution of galaxy populations and
 luminosity functions. {\it JWST} is also revealing a new physical
 picture of early galaxies, characterized by compact, clumpy star
 formation, high gas fractions, and intense radiation fields.  Despite
 much progress, many open questions remain, including the efficiency,
 intermittency, and stellar initial mass function (IMF) of early star
 formation, conditions in the interstellar medium (ISM), masses, and
 structure of the earliest galaxies, and the topology of reionization.
 Resolving these open questions will require lensing magnification,
 sensitive spectroscopy to $>5\,\mu$m, including with the Atacama
 Large Millimeter Array (ALMA), exploration to higher redshifts, and
 larger, more complete samples, helping to clarify how the first
 galaxies formed and evolved.
\end{abstract}

\section{Context and Introduction}

Understanding the origin and evolution of the universe has long been a
central goal of astronomy, and progress in this effort has accelerated
rapidly over the past three decades.  Setting the stage for current
advances in early galaxy studies were precise observations of the
cosmic microwave background with the {\it WMAP} and {\it Planck}
satellite missions. These observations provided a detailed picture of
the early universe and established the initial conditions from which
cosmic structure formed \citep{WHu2002,Planck2020_A1}.  In parallel,
dramatic improvements were made in our ability to observe the sky at
optical and near-infrared wavelengths, with the {\it Hubble Space
  Telescope} ({\it HST}) providing a view of galaxies forming as early
as $\sim$400 million years after the Big Bang
\citep{Bouwens2011,Oesch2016}.  Over the past 4 years, progress has
accelerated dramatically thanks to the {\it James Webb Space
  Telescope} and its ability to obtain sensitive,
high-spatial-resolution observations of sources out to 28 $\mu$m
\citep{GRieke2015,Jakobsen2022,Ferruit2022,Gardner2023,Doyon2023,MRieke2023}.

Following the early 20th-century discovery that distant nebulae are
island universes, longstanding questions have persisted about the
build-up and evolution of galaxies, accompanied by a wide range of
theoretical expectations. Views ranged from the idea that the
high-redshift universe might be rich in Lyman-$\alpha$ (Ly$\alpha$)
emission, the strongest ultraviolet spectral line of hydrogen,
\citep{Partridge1967} to the universe being filled with bright early
counterparts of massive ellipticals seen today \citep{Tinsley1980}.
There was also some skepticism that z$>$1 galaxies were very luminous
given null search results early on
\citep{Pritchet1994,Djorgovski1992}.

The study of galaxies at the earliest times is possible because of the
finite speed of light, which leads to long travel times for radiation
from distant sources; as this light travels toward us cosmic expansion
stretches its wavelength, so the earlier a galaxy emitted its light,
the more strongly redshifted it appears today.  The redshift, denoted
by $z$, measures this wavelength stretch: the universe has expanded by
a factor $1+z$ since the light was emitted.  Given a model for the
expansion history, redshift can therefore be translated into a
look-back time or, equivalently, the age of the universe when the
light was emitted.

\begin{figure*}
\centerline{
  \psfig{file=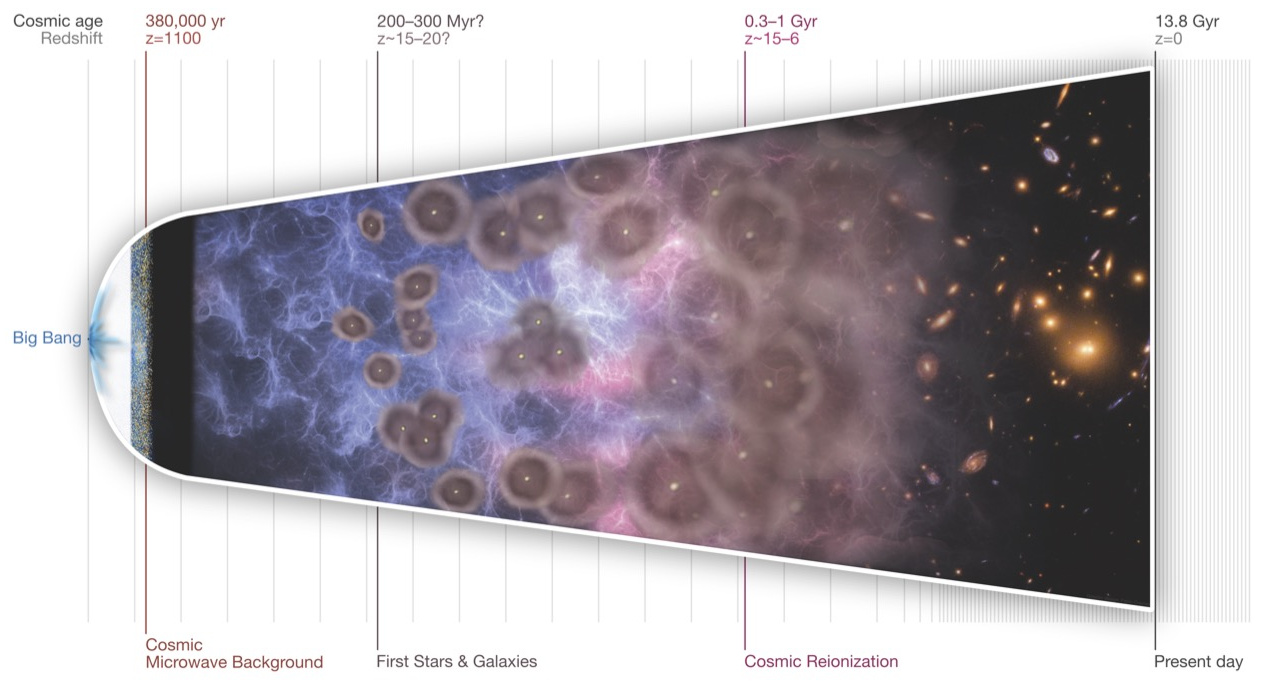,width=0.9\textwidth}}
\vspace{-0.35cm}
\caption{A schematic history of the universe from the Big Bang to the present day. Cosmic time appears on the horizontal axis, while the vertical axis shows the universe’s relative size on an approximate logarithmic scale. The figure marks the redshift of last scattering (photon decoupling from hydrogen), the emergence of the first stars and galaxies, the epoch of cosmic reionization, and the present day. Adapted from media produced for the DELPHI project and shown in Adamo et al. \citep{Adamo2025}.\label{fig:cosmic-hist}}
\end{figure*}

Searches for star-forming galaxies in the early universe now enable
the spectroscopic confirmation of systems forming when the universe was only
$\sim$300 million years old, corresponding to a redshift of $\sim$14.5
\citep{Robertson2024,Carniani2024,Naidu2026}. At this epoch, the
universe was $\sim$15$\times$ smaller in linear scale, and matter
density fluctuations were small, typically at the level of one part in
$\sim$100. This also lies at the end of the so-called cosmic “dark
ages,” the period after recombination when protons and electrons
combined into neutral hydrogen and no luminous sources had formed (Figure~\ref{fig:cosmic-hist}).  By
this time, moderately massive galaxies were assembling in significant
numbers \citep{Barkana2001,Somerville2015,Dayal2018}, and current
discoveries suggest that even earlier sources will soon be
spectroscopically confirmed
\citep{Asada2026,Kokorev2025_z16,Castellano2025,PerezGonzalez2025}. The
first stars, black holes, and galaxies likely formed at least 100–150
Myr earlier
\citep{Abel2002,Bromm2004,Glover2013,Bromm2013,Klessen2023}.

\begin{figure*}[t]
\centerline{
\psfig{file=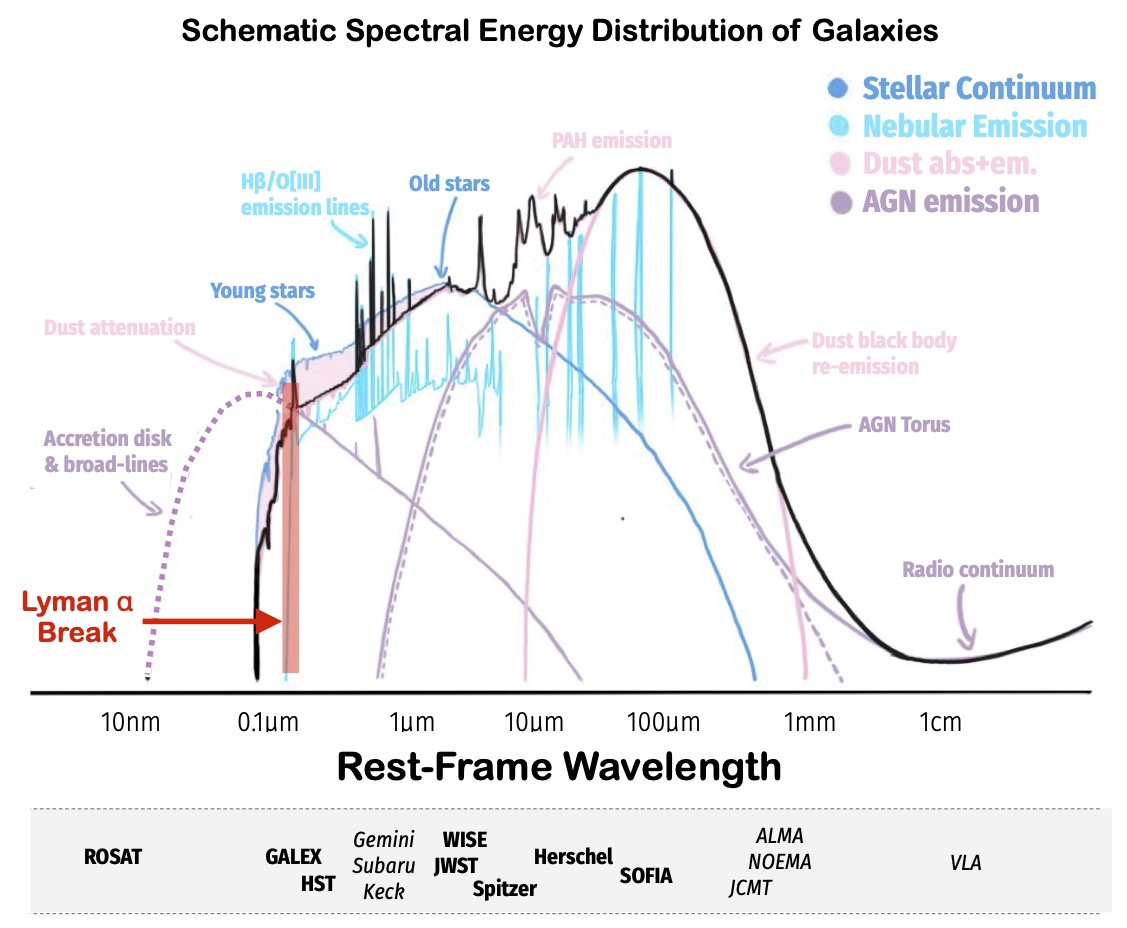,width=0.9\textwidth}}
\vspace{-0.45cm}
\caption{Illustration of the four primary sources of radiation from
  galaxies. The vertical axis shows flux output, and the horizontal
  axis indicates rest-frame wavelength. Components include stellar
  emission (\textit{blue}), nebular emission from ionized gas around
  hot stars (\textit{cyan}), dust emission and absorption
  (\textit{pink}), and emission from active galactic nuclei
  (\textit{light violet}). Nebular emission includes both strong
  emission lines--such as Ly$\alpha$, [O II]$_{3727,3729}$, [O
    III]$_{4959,5007}$, H$\beta$, and H$\alpha$ -- and a weaker
  continuum emission (see Figure~\ref{fig:nebcont}). The approximate
  position of the Lyman-$\alpha$ spectral break used to select distant
  galaxies is marked in red. The lower gray band highlights the
  wavelengths at which various observational facilities operate.
  Adapted from Iyer et al. \citep{Iyer2026}.\label{fig:toy_spectrum}}
\end{figure*}

A central component of studying distant star-forming galaxies has been
their identification in deep imaging obtained with powerful
telescopes. Observations in $\geq$3 wavelength channels enable
estimates of both the nature and redshift of these galaxies by
measuring brightness in discrete wavelength bands—effectively a
low-resolution spectrum.  These flux measurements are invaluable given
the diversity of galaxy spectral energy distributions (SEDs)
\citep{Conroy2013,Leja2017,Iyer2026} and are critical for separating
distant galaxies from nearer contaminants. Galaxies emit across a wide
range of wavelengths, reflecting contributions from several physical
components, including stars, ionized gas, dust, and active galactic
nuclei (Figure~\ref{fig:toy_spectrum}).

Beginning in the early 1990s and continuing through the era leading up
to {\it JWST}, there was substantial work focused on estimating
redshifts for large galaxy samples
\citep{Guhathakurta1990,Steidel1992,Madau1996,Steidel1996} and
validating them spectroscopically
\citep{Lilly1995,Steidel1999,Steidel2004,LeFevre2005,Vanzella2006,Barger2008,Stark2010,Pentericci2011,Newman2013,Steidel2014,LeFevre2015,Kriek2015,Bacon2017,Inami2017,Pentericci2018,Weaver2022,Harikane2022,Bouwens2022}. Identifying
distant galaxies has often been easier than identifying
intermediate-redshift sources because high-redshift galaxies exhibit a
sharp spectral break caused by the absorption of ultraviolet photons
at rest-frame $\sim$900–1216\,\AA$\,$ by neutral hydrogen
\citep{Madau1995,Songaila2004,Meiksin2006,Inoue2014}. Progress toward
identifying galaxies at higher redshift has followed advances in red
optical and increasingly sensitive near-infrared facilities, from {\it
  HST}'s reach to 1.7 $\mu$m
\citep{Thompson1998,Sirianni2005,Kimble2008} to {\it JWST}’s extension
to 5.5 $\mu$m and beyond
\citep{GRieke2015,Jakobsen2022,Ferruit2022,Gardner2023,Doyon2023,MRieke2023},
capabilities that have proven to be transformative.

The revolution enabled by {\it JWST} stems not only from its
unparalleled near-infrared imaging, which enables sensitive
multi-wavelength selection of distant galaxies, but also from
exceptionally sensitive spectroscopy that allows redshift
determinations and detailed physical characterization for large
samples.  The lack of comparable spectroscopic capability,
particularly at $>$2.4 $\mu$m, significantly limited studies during
the {\it HST} era.

The observations reviewed here address five connected questions.  When
did galaxies first form, and how rapidly did they build up in mass?
What was the overall make-up of these galaxies in terms of their
stars, gas, and chemical composition?  How quickly did the first
generations of stars enrich galaxies with elements heavier than helium
and produce dust?  When did massive black holes first appear, and how
ubiquitous were they in early star-forming galaxies?  Finally, did the
ultraviolet radiation from early galaxies drive the ionization of the
hydrogen in intergalactic medium, ending the cosmic dark ages?  These
questions organize the discussion that follows.

This review is organized as follows. \S2 describes the capabilities of
the {\it James Webb Space Telescope} and its instruments. \S3 explains
how distant-galaxy candidates are identified and how their redshifts
are confirmed spectroscopically. \S4 examines the build-up of galaxies
over cosmic time through their star-formation rates, masses, sizes,
and evolving abundance. \S5 explores the physical nature and diversity
of early galaxies, including their gas reservoirs, ionized gas,
chemical enrichment, central black holes, and the emergence of
quiescent systems. \S6 reviews progress in understanding cosmic
reionization, \S7 discusses several areas in which rapid progress
continues to be made, and \S8 provides a summary.

\section{James Webb Space Telescope: Tools for a Revolution}

Recent advances in the study of early galaxies have been driven both by
the remarkable capabilities of the telescope itself--particularly its
light-collecting capabilities--and by the powerful suite of
instruments it carries.

\subsection{Instrumentation}

The {\it James Webb Space Telescope} features four primary scientific
instruments:\\

\vspace{-0.4cm}\noindent\hspace{0.25cm}(1) \textbf{NIRCam}, a near-infrared (0.6–5.0 $\mu$m) imager and
wide-field slitless spectrograph;\\

\vspace{-0.4cm}\noindent\hspace{0.25cm}(2) \textbf{NIRSpec}, a near-infrared (0.6–5.3 $\mu$m) multi-object spectrograph with integral field unit (IFU) and fixed-slit modes;\\

\vspace{-0.4cm}\noindent\hspace{0.25cm}(3) \textbf{MIRI}, a mid-infrared (5–28 $\mu$m)
imager, spectrograph, and coronagraph; and\\

\vspace{-0.4cm}\noindent\hspace{0.25cm}(4) \textbf{NIRISS}, a near-infrared (0.6–5.0 $\mu$m) imager and slitless spectrograph.\\\vspace{-0.3cm}

\begin{figure*}[t]
\centerline{
\psfig{file=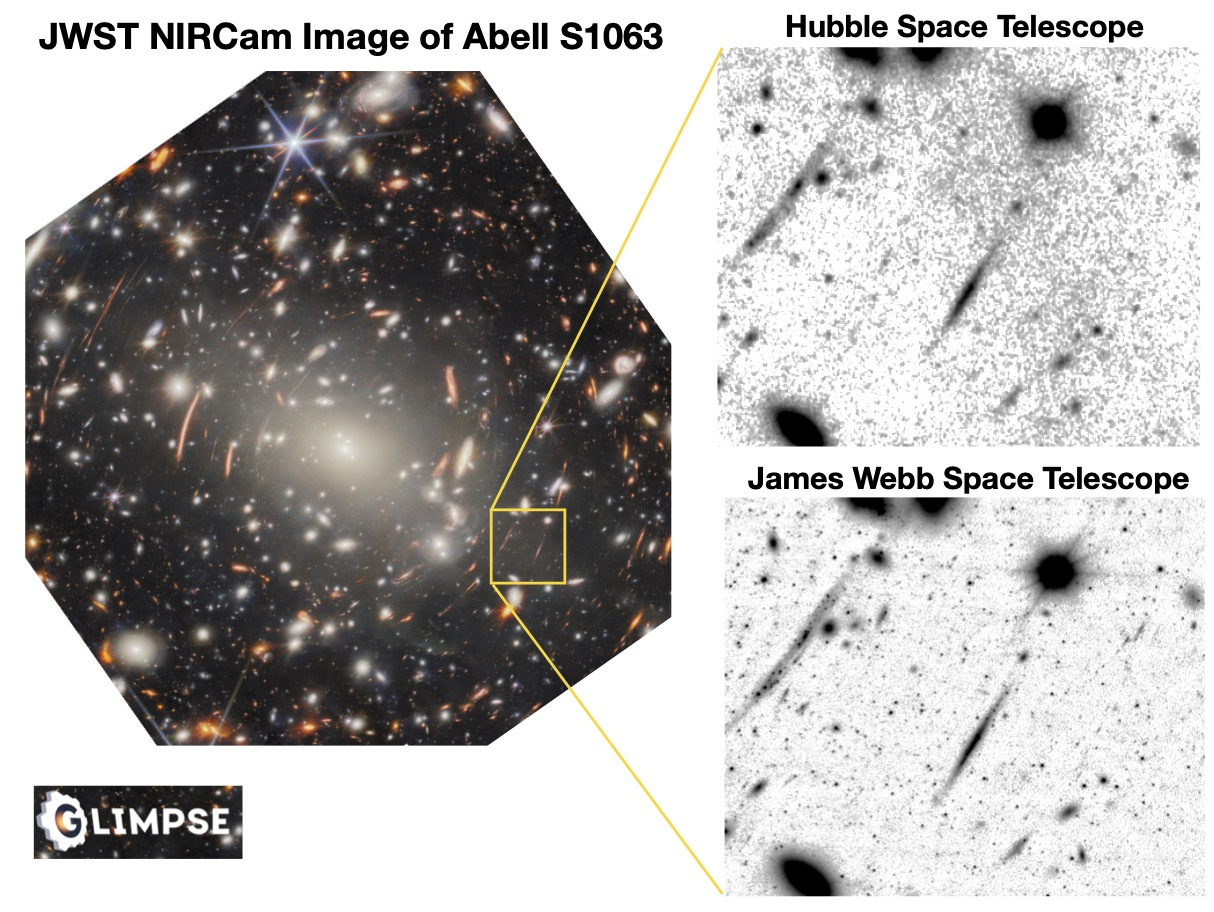,width=0.9\textwidth}}
\vspace{-0.4cm} 
\caption{(\textit{Left Panels}) False-color image of the cluster field
  Abell S1063 observed with {\it JWST} \citep{Atek2025_glimpse}, providing
  us with one of the most sensitive views of the distant
  universe. (\textit{Right panels}) Zoom-in on the same
  $24''\times22''$ region behind the cluster comparing 49 hours of {\it HST}
  imaging \citep{Lotz2017} using the F105W, F125W, F140W, F160W
  filters (0.9–1.7 $\mu$m) with 80 hours of {\it JWST}/NIRCam data using the
  F115W and F150W filters (0.9–1.7 $\mu$m). The {\it JWST} observations
  deliver a dramatic $\sim6\times$ gain in sensitivity and
  $\sim3\times$ improvement in resolution.\label{fig:deepfield}}
\vspace{-0.2cm}
\end{figure*}

Each instrument has its own observational niche for probing the
properties of distant galaxies. NIRCam provides an especially
efficient means of obtaining high–spatial-resolution (30–80 mas)
imaging over modest sky areas.  It simultaneously covers an $\sim$8
arcmin$^2$ field at both shorter (0.6–2.4 $\mu$m) and longer (2.4–5.0
$\mu$m) wavelengths \citep{MRieke2023}. Similar capabilities are
largely available with the NIRISS instrument \citep{Doyon2023}, though
over a smaller $\sim$4 arcmin$^2$ field of view (half that of NIRCam).

NIRSpec enables highly sensitive spectroscopy of faint sources in both
the nearby and distant universe \citep{Jakobsen2022,Ferruit2022}. Its
three dispersive elements (R$\sim$100, 1000, 2700) allow both
efficient detection of continuum emission and the resolution of
closely spaced spectral lines. Configurable slits---small openings
that can be selected electronically to isolate chosen targets---reduce
unwanted background light.  NIRSpec can open many of these slits
across its 2.5$\times$2.5 arcminute focal plane and obtain spectra of
$\sim$100–150 selected sources simultaneously. In
addition, the integral field unit (IFU) \citep{Boker2022} provides
spatially resolved spectroscopy, delivering $\sim$900 spectra over a
3$\arcsec\times$3$\arcsec$ region with 0.1$\arcsec$ spatial sampling.

MIRI extends JWST’s reach to wavelengths $>$5 $\mu$m
\citep{GRieke2015,GWright2015} through both imaging and single-slit
spectroscopy \citep{Kendrew2015}. These capabilities probe continuum
emission from older, lower-mass stellar populations while also
accessing key rest-frame optical emission lines used to characterize
distant galaxies.

\begin{figure*}
\centerline{
  \psfig{file=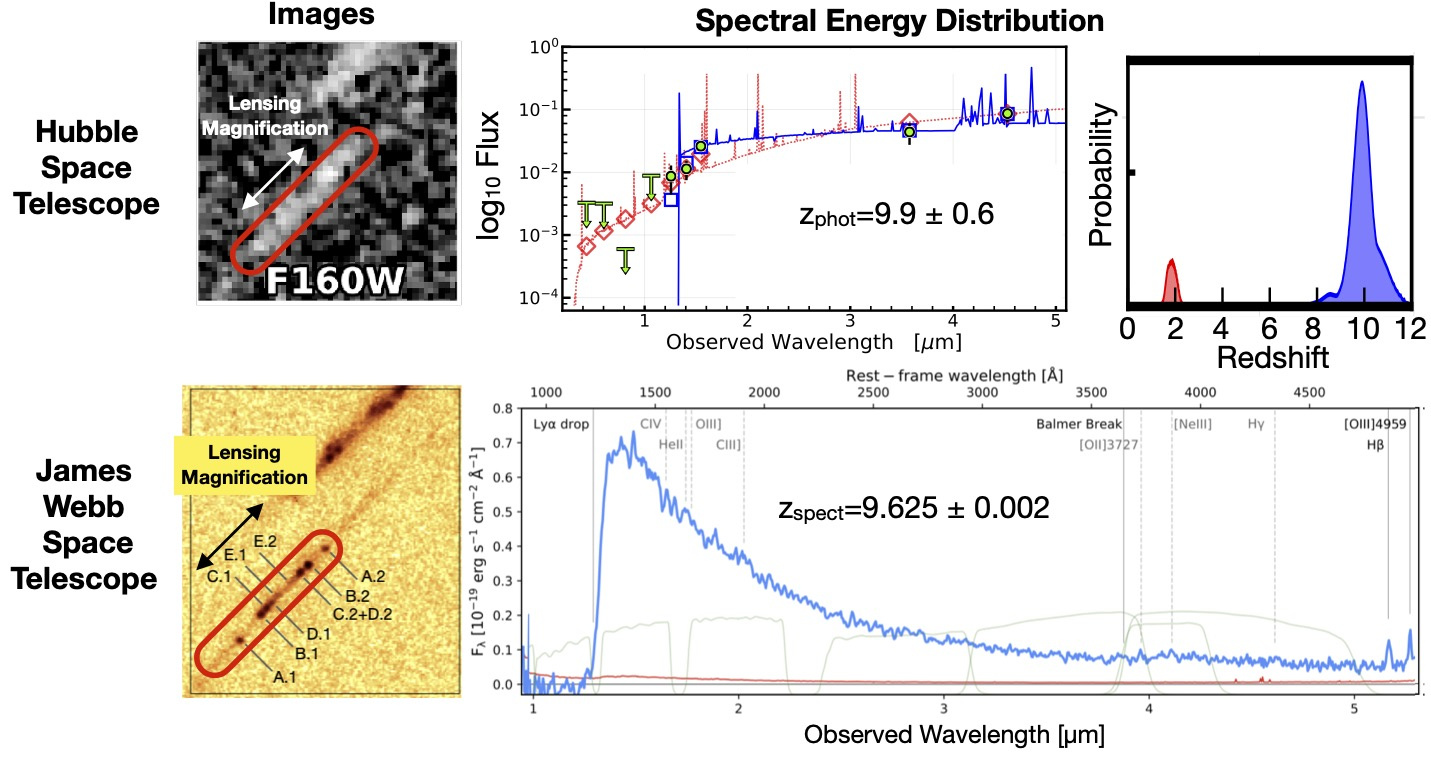,width=0.96\textwidth}}
\vspace{-0.4cm}
\caption{(\textit{Upper and lower left panels}) Images of the
  $z=9.625$ Cosmic Gems source obtained with {\it HST} \citep{Salmon2018}
  and {\it JWST} \citep{Messa2025}.  Strong gravitational lensing stretches
  the source by factors of $\sim$80–300 \citep{Messa2025}, allowing
  individual star-forming clumps to be resolved on scales of $\sim$1
  pc (parsec; one parsec is about 3.26 light-years).
  (\textit{Upper and lower right panels}) Measurements in
  several broad {\it HST} imaging bands provide a sparsely sampled
  spectral energy distribution and hence a broad redshift probability
  distribution.  By directly dispersing the galaxy's light,
  {\it JWST}/NIRSpec instead provides a continuous spectrum and a much
  more precise redshift.\label{fig:jwstdq}}
\end{figure*}

Finally, NIRCam and NIRISS provide wide-field slitless spectroscopy
\citep{Greene2017,Doyon2023}.  Rather than isolating selected sources
with slits, this mode disperses the light of every source in the
field.  It can therefore discover emission-line galaxies
serendipitously, without a prior target list.  The trade-off is that
spectra can overlap and the sky background is higher, reducing
sensitivity by $\gtrsim$5$\times$ compared to NIRSpec observations of
similar duration.

\subsection{Comparison with Previous Facilities}

The capabilities of the {\it James Webb Space Telescope} represent a
major advance relative to previous facilities. To understand why, it
is useful to compare {\it JWST} with {\it HST}, itself a revolutionary
observatory. Beyond its state-of-the-art instruments, a key advantage
is {\it JWST}’s 25 m$^2$ collecting area—about 5.5$\times$ larger than
that of {\it HST} \citep{Gardner2023}. Even assuming comparable
detectors, this means {\it JWST} can detect galaxies roughly a
magnitude fainter than {\it HST} in comparable exposure times
(Figure~\ref{fig:deepfield}). Astronomers measure brightness on a
logarithmic ``magnitude'' scale in which larger numbers denote fainter
sources and each magnitude corresponds to a factor of $\sim$2.5 in
brightness, so one magnitude deeper means detecting galaxies
$\sim$2.5$\times$ fainter.

Second, {\it JWST}’s larger primary mirror enables substantially
higher spatial resolution observations of distant galaxies.  The
angular resolution of a telescope scales roughly as $\lambda/D$, where
$\lambda$ is the observed wavelength and $D$ is the telescope
diameter.  Whereas {\it HST} achieves a diffraction limit of $\sim0.1$
arcsec at 1$\mu$m, {\it JWST} achieves a limit of $\sim$0.037 arcsec
at the same wavelength. Given the small sizes of early galaxies, this
improved resolution allows light to be measured over smaller sky
areas, reducing background noise and boosting
signal-to-noise. Consequently, {\it JWST} can probe faint
high-redshift galaxies up to 2 magnitudes (a factor of $\sim$6)
fainter than {\it HST} at
1.5 $\mu$m (Figure~\ref{fig:jwstdq}).

Third, {\it JWST} extends observations to significantly longer
infrared wavelengths than {\it HST}. While there is some overlap in
their wavelength coverage, {\it JWST} reaches out to 5.3 $\mu$m with
NIRCam and NIRSpec and to 28$\mu$m with MIRI, allowing astronomers to
detect galaxies whose light has been strongly redshifted by cosmic
expansion.

In probing light at 2–2.4 $\mu$m and at $>$2.5 $\mu$m, the closest
comparisons are with large ground-based telescopes and the 80-cm {\it
  Spitzer Space Telescope}, respectively.  In both regimes, the gains
achieved by {\it JWST} are particularly dramatic. {\it Spitzer}
achieved resolutions of $\sim$1.2 arcsec at $\sim$4 $\mu$m and
$\sim$7.2 arcsec at 24 $\mu$m, compared to $\sim$0.14 arcsec and
$\sim$0.8 arcsec for {\it JWST} at the same wavelengths. Combined with {\it
  JWST}’s larger collecting area, this improvement yields roughly a
30$\times$ gain in sensitivity across the wavelength range 3–28 $\mu$m
\citep{GRieke2015,MRieke2023}.

\section{{\it JWST} as a Redshift Machine}

\subsection{From Candidates to Confirmed Galaxies with Redshifts}

\begin{figure*}[t]
\centerline{ \psfig{file=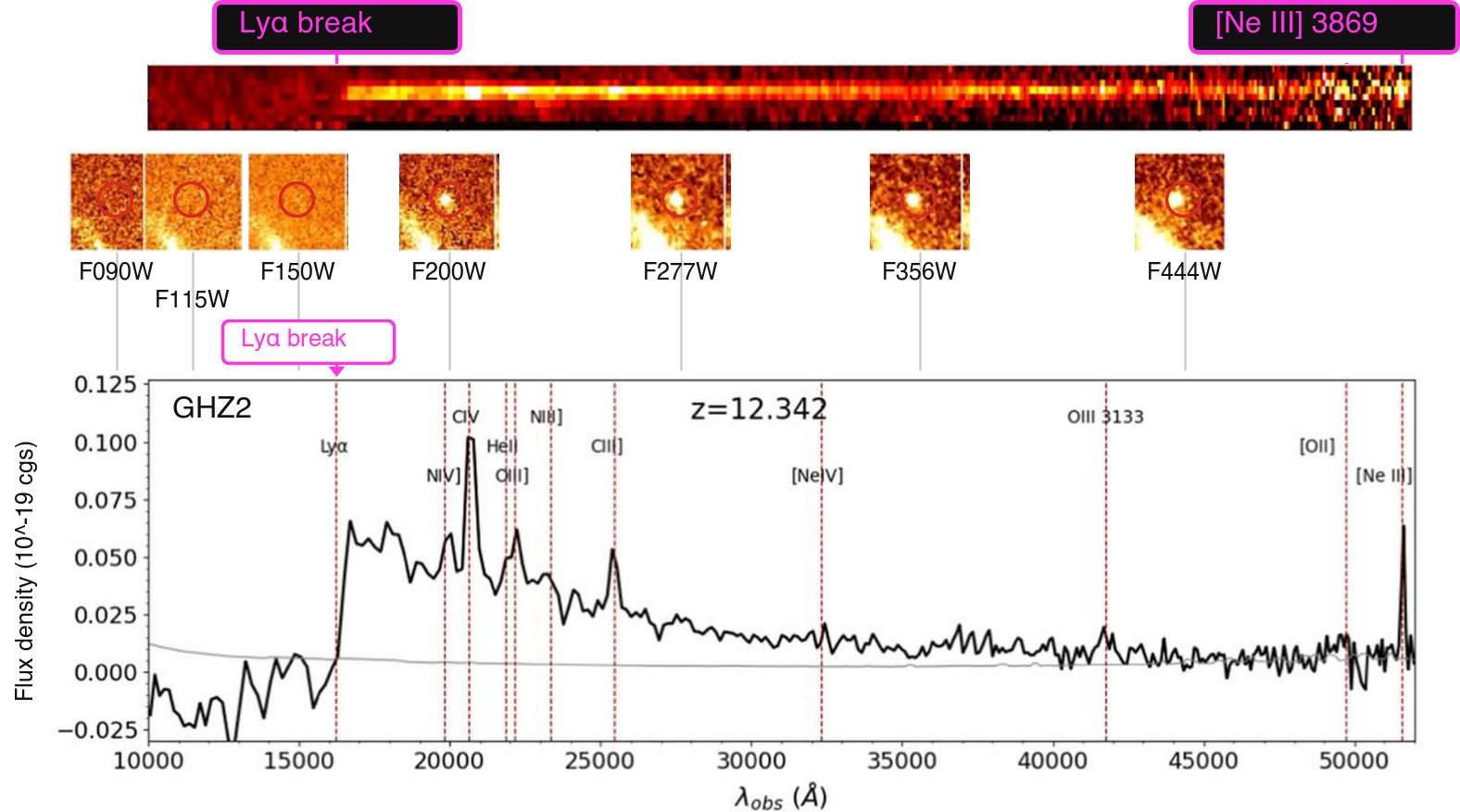,width=0.88\textwidth}}
\vspace{-0.55cm}
\caption{Illustration of the Lyman-break identification and
  spectroscopic confirmation technique, using GHZ2/GL-z12 at
  $z=12.342$ \citep{Castellano2022,Naidu2022,Castellano2024}.
  (\textit{Upper panel}) Two-dimensional {\it JWST}/NIRSpec spectrum,
  with the Lyman-$\alpha$ break and the [Ne III]$_{3869}$ emission
  line marked; wavelength runs horizontally. (\textit{Middle panel})
  NIRCam postage-stamp images from F090W to F444W: the source is
  undetected in the bluest bands (F090W--F150W) but shows clear,
  rapidly brightening continuum emission redward of the Lyman-$\alpha$
  break -- the signature used to select high-redshift galaxies.
  (\textit{Lower panel}) The full one-dimensional NIRSpec spectrum, in
  which lines from carbon, nitrogen, oxygen, and neon provide a
  precise spectroscopic redshift. \label{fig:lbg}}
\end{figure*}

To leverage {\it JWST}’s revolutionary imaging capabilities and
conduct a deep census of early galaxies, astronomers obtained highly
sensitive observations over roughly one square degree during the first
year of operations--about four times the area of the full moon. These
programs included deep imaging of SMACS0723 \citep{Pontoppidan2022}
and Abell 2744 \citep{Bezanson2024}, several massive galaxy clusters
\citep{Windhorst2023,Sarrouh2026}, an even deeper view of the Hubble
Ultra Deep Field \citep{Beckwith2006,Eisenstein2026_JADES}, and
multiple legacy {\it HST} survey fields
\citep{Casey2023_COSMOSWeb,Williams2025,Finkelstein2025_CEERS,Windhorst2023,Bagley2024,Donnan2024}. The
surveys covered wavelengths from 0.8 $\mu$m to 28 $\mu$m, enabling
measurements of stellar emission, ionized gas, hot dust, and AGN
activity, while delivering dramatic gains beyond {\it Hubble}’s reach
at $>$1.6 $\mu$m \citep{Rieke2024,Alberts2024}.  Early analyses of the
first {\it JWST} images quickly revealed several extremely bright
star-forming galaxies in the early universe, identified from sharp
spectral breaks in multi-band imaging. These included robust
candidates out to $z\sim13$ (Figure~\ref{fig:lbg}), an apparently
luminous system at $z\sim16$, and several more tentative high-redshift
candidates
\citep{Castellano2022,Naidu2022,Atek2023,Finkelstein2022,Donnan2023,
  Harikane2023,Adams2023,Bouwens2023,Morishita2023}.

The first clear demonstration of {\it JWST}'s spectroscopic power came
from observations of gravitationally lensed galaxies at
$z\sim6.5$--8.5
\citep{Schaerer2022,ArellanoCordova2022,Taylor2022,Trump2023,
  Tacchella2023,Brinchmann2023,Heintz2023,Curti2023,Fujimoto2024_z8p5},
corresponding to approximately 600 Myr after the Big Bang.
Spectroscopic confirmations at still higher redshifts soon followed,
with multiple sources confirmed at $z>8.5$
\citep{RobertsBorsani2023,WilliamsH2023,Nakajima2023,Tang2023,
  Fujimoto2023,ArrabalHaro2023Nat,CurtisLake2023,Wang2023,Boyett2024}.
These included galaxies extending to $z=13.2$, only about 300 Myr
after the Big Bang \citep{CurtisLake2023}. Remarkably, within five
months of beginning science operations, {\it JWST} had both identified
galaxies farther back in cosmic time than ever before and
spectroscopically established the distances of several record-breaking
sources.

The distance record has since been broken twice. A bright star-forming
galaxy at $z=14.2$ \citep{Carniani2024} was identified in a deep
{\it JWST}/NIRCam field obtained by a GTO team
\citep{Eisenstein2025}, followed by a galaxy at $z=14.44$ in the
PRIMER-COSMOS field \citep{Naidu2026}. These confirmations provide
glimpses of galaxies at still earlier times and help constrain their
build-up and diversity. They also show that carefully selected
photometric candidates at $7<z<14$ are generally reliable, although
occasional lower-redshift contaminants demonstrate the continuing
importance of spectroscopy.

\begin{figure*}[t]
\centerline{
  \psfig{file=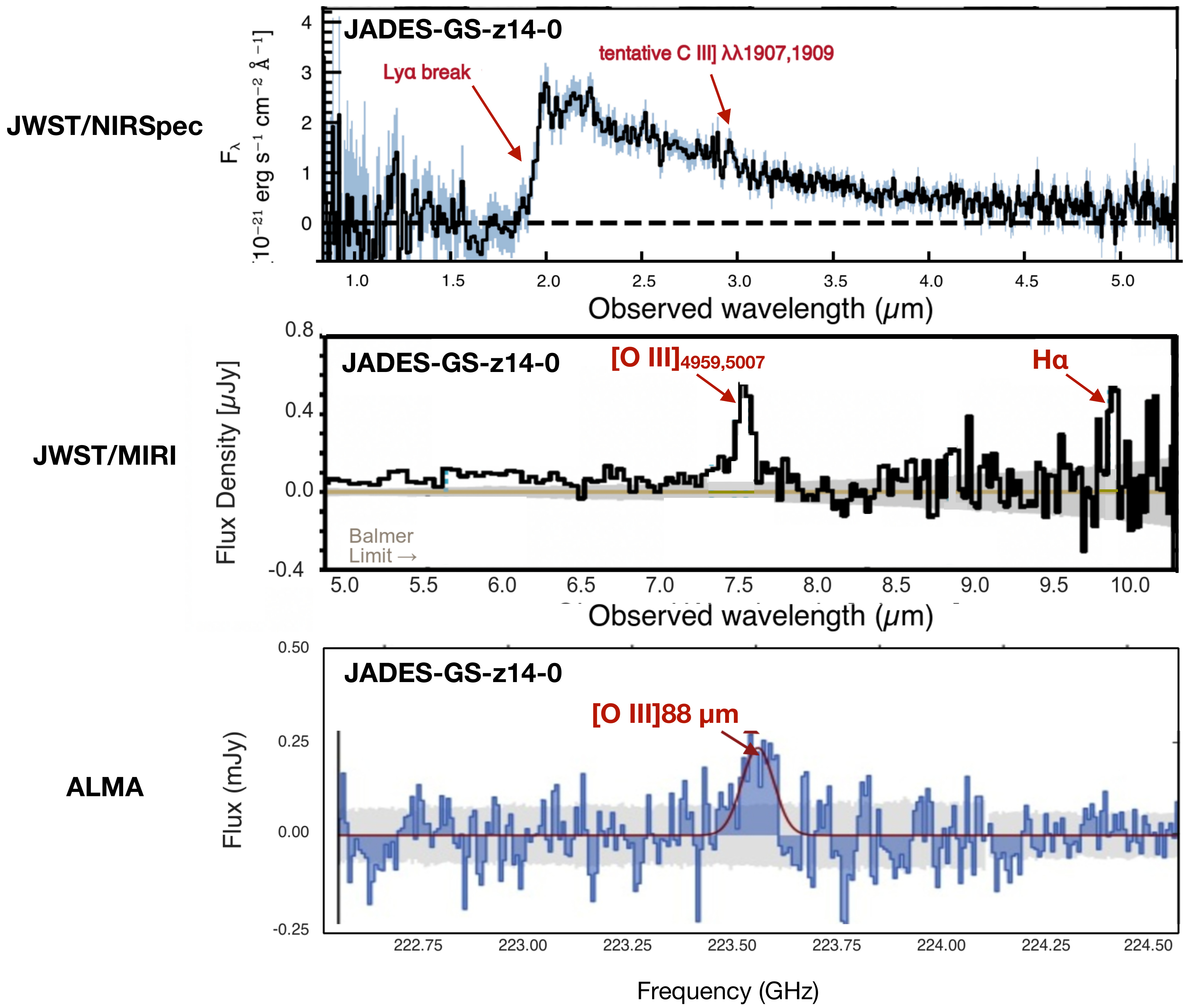,width=0.95\textwidth}}
\vspace{-0.3cm}
\caption{Independent spectroscopic confirmations of JADES-GS-z14-0 at
  $z=14.18$. (\textit{Upper}) {\it JWST}/NIRSpec low-resolution,
  extracted one-dimensional spectrum, with a strong Lyman-$\alpha$
  break present and a tentative ($\sim$3$\sigma$) CIII]$_{1907,1909}$
line indicated \citep{Carniani2024}.  (\textit{Middle}) {\it
  JWST}/MIRI low-resolution, extracted one-dimensional spectrum, with
[O III]$_{4959,5007}$ detected at $\sim$11$\sigma$ and H$\alpha$ at
$\sim$4$\sigma$ \citep{Helton2026}. (\textit{Lower}) Independent ALMA
detection of the far-infrared [O III]$_{88\,\mu\mathrm{m}}$ cooling
line at the same redshift, achieved in just 3 hours of integration
\citep{Schouws2025,Carniani2025}.  Measurement of the same redshift
with {\it JWST}/NIRSpec, {\it JWST}/MIRI, and ALMA illustrates the
robustness of {\it JWST}-era distance
measurements.\label{fig:redshiftmeth}}
\end{figure*}

Whereas Figure~\ref{fig:lbg} illustrates how such galaxies are
selected and confirmed, a natural question is whether these redshifts
can be trusted.  Redshifts for many of the most distant galaxies have
now been measured with multiple instruments that trace line emission
across the electromagnetic spectrum, and agreement between independent
facilities provides a particularly compelling demonstration of {\it
  JWST}'s reliability.  A striking example is JADES-GS-z14-0, which
held the spectroscopic distance record at $z=14.1793$ until 2025
\citep{Robertson2024,Carniani2024}.  Its redshift, first established
with NIRSpec, has since been confirmed independently with {\it
  JWST}/MIRI, whose ultra-deep rest-frame optical spectrum detects [O
  III]$_{4959,5007}$ at $\sim$11$\sigma$ and H$\alpha$ at
$\sim$4$\sigma$ \citep{Helton2026}, and with ALMA, which detected the
far-infrared [O III]$_{88\,\mu\mathrm{m}}$ cooling line at the same
redshift in just 3 hours of integration
\citep{Schouws2025,Carniani2025} (Figure~\ref{fig:redshiftmeth}).
Similar multi-instrument redshift confirmations have been obtained for
GHZ2/GL-z12 at $z=12.33$
\citep{Castellano2024,Zavala2024,Zavala2025} and UDFj-39546284
(GS-z11-0) at $z=11.122$
\citep{CurtisLake2023,Hainline2024,Witstok2026}.  The number of such
independently confirmed systems should increase as ongoing MIRI and
ALMA programs are completed.

Spectroscopy also guards
against occasional contaminants. One early $z\sim16$ candidate,
CEERS-93316 \citep{Donnan2023}, was shown by NIRSpec to be a dusty
star-forming galaxy at $z=4.91$ whose strong emission lines mimicked a
high-redshift spectral break \citep{ArrabalHaro2023Nat}. Such
contaminants
\citep{Stanway2003,Dunlop2013,Bouwens2015,Finkelstein2016,Juodzbalis2023,Kocevski2025,Hviding2025}
appear to be rare in high-quality {\it JWST} selections, and
spectroscopic follow-up has enabled astronomers to construct large,
robust samples of galaxies in the early universe.

\begin{figure*}[t]
\centerline{
  \psfig{file=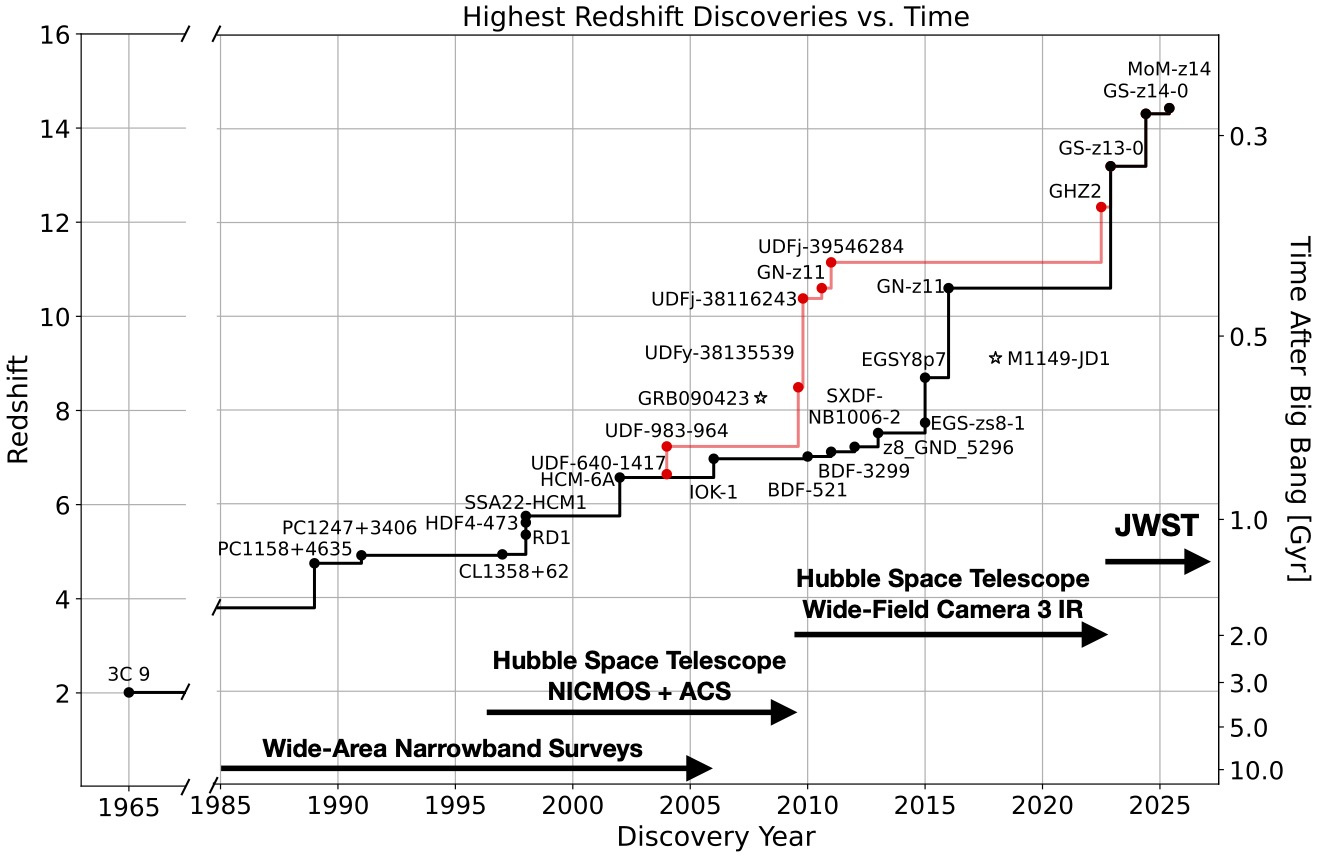,width=0.98\textwidth}}
\vspace{-0.55cm}
\caption{The most distant spectroscopically confirmed galaxy as a
  function of discovery year. Red circles and a connecting line denote
  sources initially identified as record holders based on photometry
  and later confirmed with {\it JWST} spectroscopy—particularly
  relevant from 2004–2022, when spectroscopic capabilities lagged
  photometric surveys. Record-breaking source IDs are labeled. Also
  shown are two notable high-redshift contenders: a GRB at $z=8.2$
  \citep{Tanvir2009,Salvaterra2009} and an [O III]-detected galaxy at
  $z=9.1$ \citep{Hashimoto2018}. Arrows along the bottom mark the
  start of operations for key techniques, instruments, and
  telescopes. Since beginning science operations, {\it JWST} has
  dramatically extended the redshift frontier to earlier cosmic
  times. A source first identified photometrically does not receive a
  second point if its later spectroscopic confirmation did not
  establish a new distance record at the date of confirmation; this is
  why UDFj-39546284 has no separate black point.
  \label{fig:highestz}}
\end{figure*}

\subsection{From Distance Records to Statistical Samples}

The discovery of galaxies at record-breaking distances is just one
outcome of {\it JWST}’s dramatic advances in spectroscopy, which have
enabled confirmation of thousands of sources at $z>6$, primarily with
NIRSpec and the NIRCam grism. Before {\it JWST}, the dominant
technique for securing redshifts for sources was through detection of
the Ly$\alpha$ emission from galaxies
\citep{Hu2002,Bunker2003,Dickinson2004,Vanzella2006,Iye2006,Stark2010,Vanzella2011,Ono2012,Finkelstein2013,Oesch2015,Zitrin2015,Inami2017},
which powered most extensions of the distance record.  Breakthroughs
also occurred via far-infrared [O III]$ _{88\,\mu\mathrm{m}}$
detections with ALMA in exceptional cases \citep{Hashimoto2018} and
through detection of the Ly$\alpha$ break using {\it HST} slitless
grism spectroscopy \citep{Oesch2016}.

To highlight {\it JWST}’s impact, Figure~\ref{fig:highestz} and
Table~\ref{tab:zrecord} trace the most distant known galaxy by year,
from the discovery of a $z=2.01$ QSO \citep{Schmidt1965} to the
current record at $z=14.44$ \citep{Naidu2026}. The distance frontier
advanced from $z\sim4$ in the mid-1980s to $z\sim7$ by 2011
\citep{Vanzella2011}.  This was followed by the discovery of
Ly$\alpha$ emitters at $z=7.215$, 7.51, 7.73 and 8.68
\citep{Ono2012,Finkelstein2013,Oesch2015,Zitrin2015} and by the bright
continuum-break galaxy GN-z11 at $z\sim11.1$ \citep{Oesch2016}.  After
the start of science operations in 2022, {\it JWST} rapidly extended
the record to $z=13.2$ \citep{CurtisLake2023}, very close to the
current record holder MoM-z14 at $z=14.44$ \citep{Naidu2026}.

\begin{table}[!t]\small
\centering
\begin{tabular}{ccccc}
  {\bf Date} & {\bf $z_{spec}$} & {\bf $z_{JWST,spec}$} & {\bf Source ID} & {\bf Technique} \\ \hline
  \hline
  1965 & 2.01 & --- & 3C 9 \citep{Schmidt1965} & QSO \\
  1970 & 2.878 & --- & 4C 05.34 \citep{Lynds1970} & QSO \\
  1973 & 3.40 & --- & OH 471 \citep{Carswell1973} & QSO \\
  1973 & 3.53 & --- & OQ 172 \citep{Wampler1973} & QSO \\  
  1982 & 3.78 & --- & PKS2000-330 \citep{Peterson1982} & QSO \\
  1989 & 4.73 & --- & PC1158+4635 \citep{Schneider1989} & QSO \\
  1991 & 4.897 & --- & PC1247+3406 \citep{Schneider1991} & QSO \\
  1997 & 4.92 & --- & CL1358+62 \citep{Franx1997} & LBG \\
  1998 & 5.34 & --- & 0140+326 RD1 \citep{Dey1998} & --- \\
  1998 & 5.60 & --- & HDF4-473.0 \citep{Weymann1998} & LBG \\
  1998 & 5.74 & --- & SSA22-HCM1 \citep{Hu1999} & LAE \\
  2001 & 6.28 & --- & J103027.10+052455.0 \citep{Fan2001} & QSO \\
  2002 & 6.56 & --- & HCM-6A \citep{Hu2002} & LAE \\
  2004 & & 7.234 \citep{Oesch2023} & UDF-983-964 \citep{Bouwens2004} & LBG \\
  2006 & 6.96 & --- & IOK-1 \citep{Iye2006} & LAE \\
  2009 & 8.23$^{\dagger}$ & --- & GRB 090423 \citep{Tanvir2009,Salvaterra2009} & GRB \\
  2009 & & 8.485 \citep{Bunker2024_specz} & UDFy-38135539 \citep{Bouwens2010} & LBG \\
  2009 & & 10.38 \citep{CurtisLake2023} & UDFj-38116243 \citep{Bouwens2011} & LBG \\
  2010 & & 10.60 \citep{Bunker2023_gnz11} & GN-z11 \citep{Bouwens2010,Oesch2014} & LBG \\
  2011 & & 11.122 \citep{CurtisLake2023,Hainline2024,Witstok2026} & UDFj‑39546284 \citep{Bouwens2011,Ellis2013} & LBG \\

  2011 & 7.008 & & BDF-521 \citep{Vanzella2011} & LBG \\
  2011 & 7.109 & & BDF-3299 \citep{Vanzella2011} & LBG \\
  2012 & 7.215 & & GN-108036 \citep{Ono2012} & LBG \\
  2013 & 7.51 & & z8\_GND\_5296 \citep{Finkelstein2013} & LBG \\
  2015 & 7.73 & & EGS-zs8-1 \citep{Oesch2015} & LBG \\
  2015 & 8.69 & & EGSY8p7 \citep{Zitrin2015} & LBG \\
  2016 & 10.60 \citep{Bunker2023_gnz11} & & GN-z11 \citep{Oesch2016} & LBG \\
  2018 & 9.1096$^\ddagger$ & & MACS1149-JD \citep{Hashimoto2018} & LBG \\
  2022 & & 12.33 \citep{Castellano2024} & GHZ2/GL-z12 \citep{Castellano2022,Naidu2022} & LBG \\
  2022 & 13.20 & & GS-z13-0 \citep{CurtisLake2023} & LBG \\
  2024 & 14.179 \citep{Schouws2025,Carniani2025}& & GS-z14-0 \citep{Robertson2024,Carniani2024}& LBG \\
  2025 & 14.44  & & MoM-z14 \citep{Naidu2026} & LBG\\
  \hline
\end{tabular}
    \vspace{0.05cm}
    \caption{Highest-redshift sources known as a function of discovery date. Two classes of sources are shown:\\
(1) those for which a newly confirmed redshift ($z_{spec}$) exceeded the previous distance record, and\\
(2) those that, based on current spectroscopic measurements ($z_{spec}$ or $z_{JWST,spec}$), appear to have been the most distant sources known at the time of discovery.  GN-z11, UDFj-39546284, and GHZ2/GL-z12 are examples of such sources which were identified $>$12 months prior to their spectroscopic confirmation by {\it JWST}.  LBG, GRB, and QSO denote Lyman Break Galaxy, Gamma-Ray Burst, and Quasi-Stellar Object (quasar).\\
GRB 090423 is marked with $\dagger$ to indicate that the detected source was the gamma-ray burst rather than its host galaxy. MACS1149-JD is marked with $\ddagger$ to denote its status as the highest-redshift source in 2018 based on an emission-line detection.}
    \label{tab:zrecord}
\end{table}

\begin{figure*}[t]
\centerline{
  \psfig{file=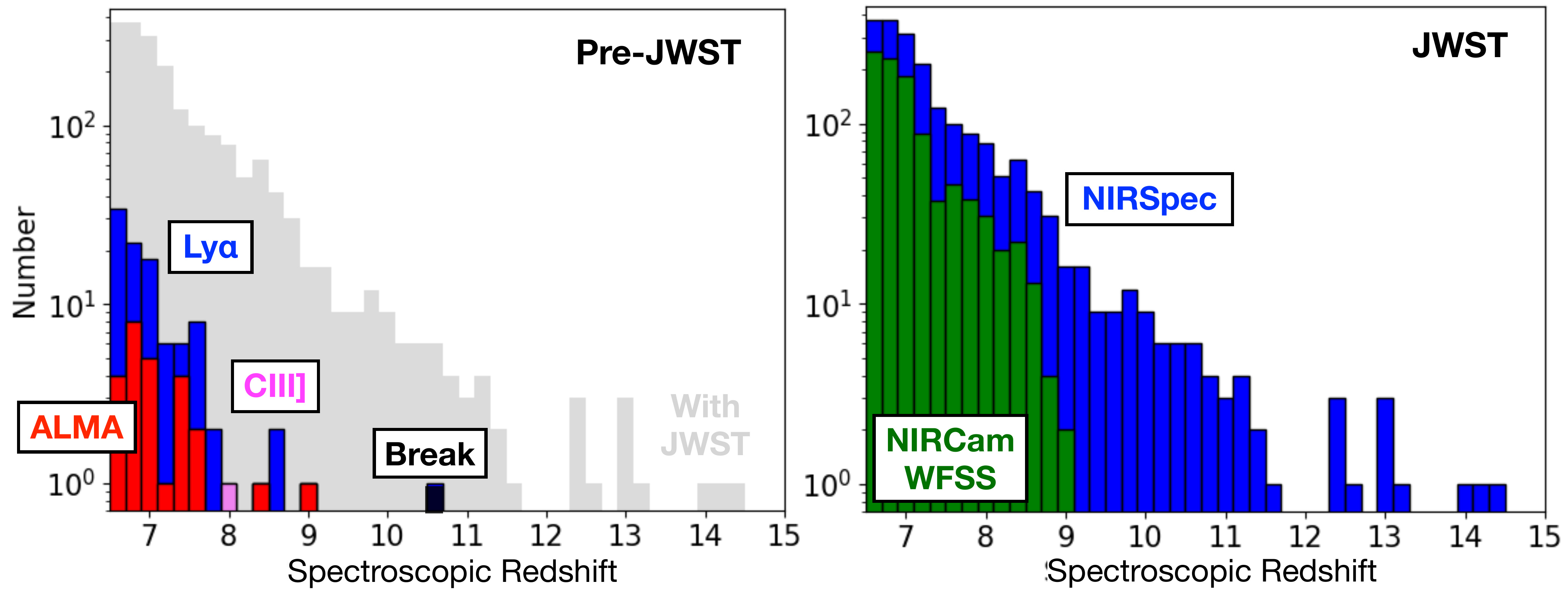,width=0.99\textwidth}}
\vspace{-0.5cm}
\caption{Histogram of secure redshift measurements prior to the start
  of {\it JWST} operations (\textit{left panel}) and from {\it JWST}
  observations to date (\textit{right panel}). Spectroscopic redshift
  determinations based on Ly$\alpha$, far-infrared lines ([C
    II]$_{158}$ or [O III]$_{88}$), C III]$_{1907,1909}$, and spectral
breaks are indicated with blue, red, magenta, and black shading,
respectively.  The grey shaded region indicates the current numbers
from {\it JWST}.  Redshift determinations from the NIRCam grism are
shown in green, while those from NIRSpec are shown in blue.  Thanks to
{\it JWST}, the number of spectroscopic redshifts available for
galaxies at z$>$6.5 and z$>$9 has increased by factors of 12 and 50,
respectively.\label{fig:jwst-z-machine}}
\end{figure*}

{\it JWST}’s impact on redshift measurements is illustrated in
Figure~\ref{fig:jwst-z-machine}, which compares the number of known
$z$$>$6 redshifts before (\textit{left}) and after (\textit{right})
{\it JWST}. Counts are grouped by discovery instrument: while NIRSpec
has delivered hundreds of new redshifts
\citep{Bunker2024_specz,RobertsBorsani2024,Heintz2025_PRIMAL,CurtisLake2026,Scholtz2026,RobertsBorsani2026,Napolitano2025,Kokorev2025,Donnan2025,Donnan2026,Rodighiero2026,Donnan2026_capers},
wide-field slitless spectroscopy with the NIRCam grism has been
equally transformative, providing thousands of redshifts
\citep{Oesch2023,HerardDemanche2025,Helton2024,Sun2024,Meyer2024,Meyer2025,Sun2025_SAPPHIRES,CoveloPaz2025,Wang2026_ASPIRE,Lin2026_grism,Sun2026JOF}.

\section{Galaxy Growth in the Early Universe}

Beyond determining when galaxies in the early universe emitted their
light and how bright they are, {\it JWST} enables us to trace their
growth with cosmic time: their star-formation rates, their stellar and
dynamical masses, their physical sizes, and their dust content.  We
then turn from individual galaxies to the population as a whole,
through the evolving luminosity and mass functions and the cosmic
star-formation rate density.

\subsection{Galaxy Star Formation Rates\label{sec:sfr}}

One important measure of galaxy growth is the star-formation rate,
which quantifies how fast the cool gas in galaxies is converted into
stars.  The star formation rate together with the duration of star
formation can provide us with an estimate of the total mass in stars a
galaxy has built up.  One simple estimate follows directly from the
brightness of galaxies in the rest-$UV$, from $\sim$1216\AA$\,$to
3600\AA.  Because this light is dominated by short-lived massive stars
($<100$ Myr), it provides a straightforward measure of recent star
formation. For the brightest known galaxies at $z>10$, inferred SFRs
typically reach $\sim10$–20 $M_{\odot}\,\mathrm{yr}^{-1}$.

Of course, rest-$UV$ emission is highly susceptible to absorption and
scattering by dust found within galaxies.  Dust consists of small
solid grains containing elements produced by stars.  The absorbed
energy heats dust grains, which then re-radiate in the far-IR, with a
peak wavelength set by the dust temperature and energy output matching
that injected by the stars themselves.  Because such elements—and
subsequently dust grains—must form and disperse through stellar
evolution, detecting dust in an early galaxy shows that chemical
enrichment had already proceeded rapidly in the early universe
(\S\ref{sec:metals}).

While some constraints on the dust content can be inferred from the
rest-$UV$ spectral shapes and hydrogen line ratios
\citep{Sandles2024}, the most direct measurements come from
far-infrared observations with facilities such as the Atacama Large
Millimeter Array (ALMA) or single-dish telescopes \citep{Smit2026}.
ALMA has detected no significant far-IR emission from the brightest
$z\geq10$ galaxies, either individually (e.g., GHZ2 and JADES-GS-z14-0
\citep{Yoon2023,Bakx2023,Popping2023,Schouws2025,Carniani2025}) or in
stacks of bright $z>8$ galaxies \citep{Bakx2026}.

\begin{figure*}[t]
  \vspace{-0.4cm}
\centerline{
\psfig{file=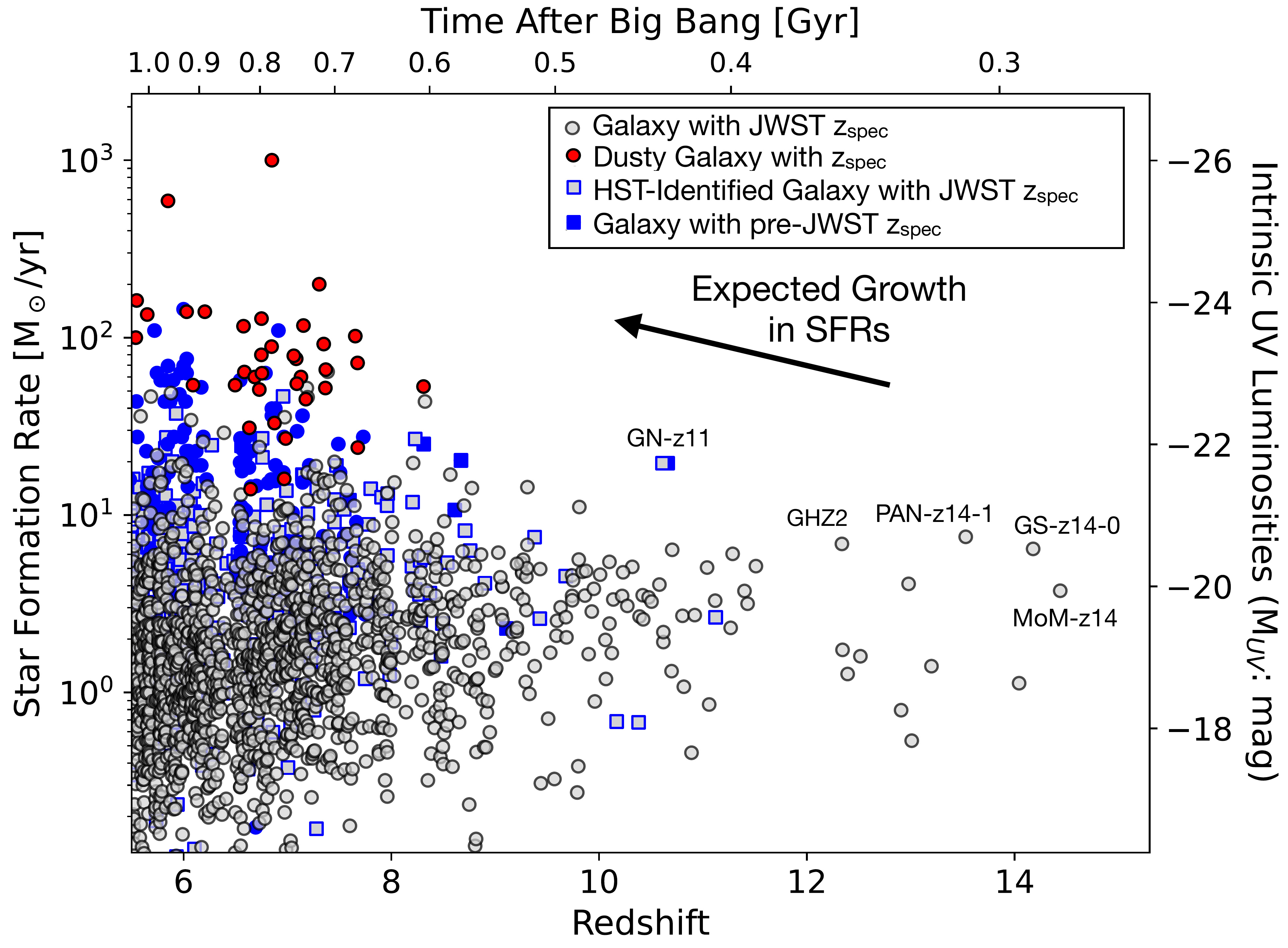,width=0.98\textwidth}}
\vspace{-0.55cm}
\caption{Estimated star-formation rate (SFR) as a function of redshift
  for spectroscopically-confirmed galaxies, with a few well-known
  $z>10$ galaxies labelled.  SFRs are derived from rest-$UV$
  luminosities and, where available, far-infrared measurements. Solid
  blue squares and gray circles indicate sources spectroscopically
  confirmed prior to and with {\it JWST}, respectively.  Galaxies
  detected in the far-IR, implying substantial dust-obscured star
  formation, are shaded red.  Sources first identified in {\it HST}
  data are shown as squares with blue boundaries.  The upper axis
  shows the corresponding cosmic age at the time of emission, and the
  right axis gives rest-UV luminosity in AB magnitude
  units.\label{fig:zsfr}}
\end{figure*}

Figure~\ref{fig:zsfr} shows the apparent star-formation rate (SFR) as
a function of redshift for current spectroscopically confirmed $z>6$
galaxies.  For uniformity, SFRs are derived from measured UV
luminosities (\textit{right vertical axis}), with additional obscured
star formation included when far-infrared continuum data are available
\citep{Laporte2017,Hashimoto2019,Tamura2019,Harikane2020,Schouws2022,Inami2022,vanleeuwen2025}. The
figure shows a clear excess of high-SFR galaxies at $z<8$ compared to
$z>8$, consistent with exponential growth in galaxy masses and SFRs
over cosmic time.

\subsection{Galaxy Masses and Sizes\label{sec:mass}}

Galaxy growth can also be traced through several complementary mass
estimates: stellar mass, dynamical mass, and the mass of the
surrounding dark-matter halo.  Stellar masses, the most widely used
estimates, are obtained by fitting galaxy photometry with population
models that describe the star-formation history and stellar
evolution. Photometric samples provide provisional estimates, while
more robust values come from spectroscopically confirmed
sources. These typically have masses of $10^8$--$10^9\,M_{\odot}$
\citep{Finkelstein2022,ArrabalHaro2023Nat,Castellano2024,Robertson2024,
  Naidu2026,Kokorev2025,Donnan2025,Donnan2026}. At their current SFRs,
these masses require star formation over $\gtrsim100$ Myr, placing its
onset at $z\gtrsim17$ \citep{Carniani2024}. If star formation
increased exponentially, formation may have begun by $z\gtrsim20$
\citep{Carniani2024}. Existing deep and forthcoming MIRI imaging and
spectroscopy at longer wavelengths will further improve these mass
estimates \citep{Ostlin2025,AlvarezMarquez2026,Song2023}.

Galaxy masses can also be estimated from the motions of ionized gas,
measured through spatial variations in the Doppler shifts of emission
lines.  These estimates may be more robust because of their direct
connection to gravity and kinematics.  Remarkably, constraints are
available as far back as $z=14.2$ for JADES-GS-z14-0 based on the
detection of the [O III]$_{88}$ \citep{Schouws2025,Carniani2025} line
with ALMA. From its line width, Carniani et al. \citep{Carniani2025}
and Schouws et al. \citep{Schouws2025} derive dynamical masses of
$10^{8.7\pm0.2}$ and $10^{8.9\pm0.2}\,M_{\odot}$, respectively. A
resolved [O III]$_{88}$ velocity gradient suggests ordered
motion. Such analyses provide direct tests of stellar-mass estimates
and constraints on other mass components, including gas
\citep{Carniani2025,Schouws2025_cii,Scholtz2025,Heintz2025_z14}.

Estimates of the masses of the collapsed dark matter in which galaxies
form help tie galaxy formation to matter fluctuations in the early
universe.  One standard method for estimating the mass of these halos
is from the measured clustering of galaxies.  Thus far, clustering
indicates dark matter halo masses of $\sim$10$^{10.5}$ $M_{\odot}$,
but current statistics are still modest
\citep{Dalmasso2024,Weibel2026_cluster,Paquereau2025}.

\begin{figure*}
\centerline{
  \psfig{file=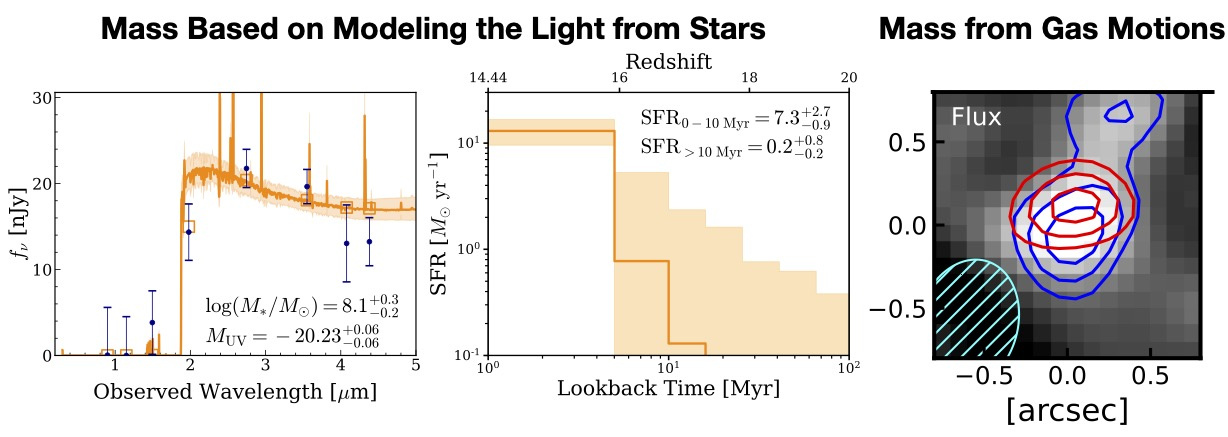,width=0.98\textwidth}}
\vspace{-0.53cm}
\caption{Constraints on the masses of galaxies within the first 300
  Myr of cosmic history. (\textit{Left}) Model of the NIRCam fluxes of
  MoM-z14 (in nJy) based on stellar population modeling
  \citep{Naidu2026}. (\textit{Center}) Reconstructed star-formation
  history of MoM-z14 (in $M_{\odot},\mathrm{yr}^{-1}$)
  \citep{Naidu2026}. (\textit{Right}) Spatial distribution of gas
  velocities (toward and away from the line of sight) in JADES-GS-z14-0 at
  $z=14.18$ \citep{Scholtz2025}. These kinematic measurements enable
  estimates of the galaxy’s dynamical mass, since greater total mass
  produces larger circular and random motions.\label{fig:mass}}
\end{figure*}

Besides mass, another measure of early galaxy growth involves the
physical sizes of these sources.  New results from {\it JWST} indicate
$z>7$ galaxies \citep{Yang2022,Ono2023} are $\sim$$2$–$3\times$
smaller than galaxies of similar mass 1–2 Gyr later in cosmic time,
extending the trend towards smaller, denser systems at early times
\citep{Ferguson2004,Bouwens2004_size,Oesch2010,Shibuya2015}.

\subsection{Emergence of Increasingly Luminous Galaxies in the Early Universe}

Beyond individual galaxies, tracing the growth of the galaxy
population as a whole lets us assess its collective impact on the
cosmic densities of stars and metals and on the ionization state of
the intergalactic medium. It also links galaxy evolution to the
environment in which galaxies form and helps identify processes that
regulate or suppress star formation.

\begin{figure*}[t]
\centerline{
\psfig{file=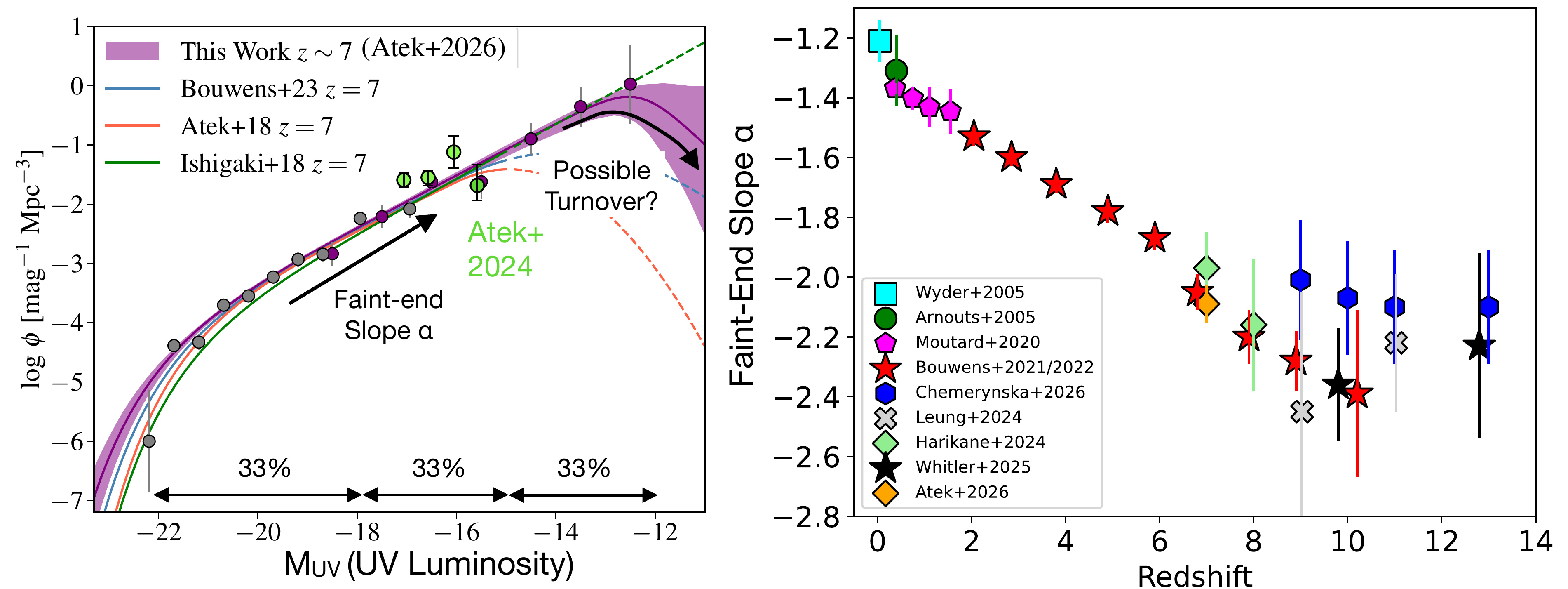,width=0.99\textwidth}}
\vspace{-0.4cm}
\caption{(\textit{Left panel}) Constraints on the abundance
  \citep{Atek2026} of low-luminosity star-forming galaxies at $z\sim7$
  (\textit{purple circles and shaded region}) from the GLIMPSE {\it
    JWST}/NIRCam program \citep{Atek2025_glimpse}. Gravitational
  lensing by the foreground cluster Abell S1063 ($z=0.348$) enables
  high-S/N measurements of faint galaxies. Also shown are {\it
    JWST}/NIRSpec results behind Abell 2744 \citep{Atek2024} and
  pre-{\it JWST} results from {\it HST} Frontier Fields
  \citep{Atek2018,Ishigaki2018,Bouwens2022_hff}. The shaded violet
  region shows current constraints \citep{Atek2026} on a possible
  turn-over below $M_{\mathrm{UV}}\sim-12.5$ mag. (\textit{Right
    panel}) Representative constraints on the faint-end slope $\alpha$
  of the rest-frame UV luminosity function
  \citep{Wyder2005,Arnouts2005,Moutard2020,Bouwens2022_hff,Leung2023,Chemerynska2026,Harikane2024,Whitler2025}. The
  slope shows monotonic flattening from early to later times, broadly
  tracking the dark matter halo mass function. For $\alpha\approx-2$,
  roughly one-third of $UV$ photons originate from galaxies brighter
  than $-18$ mag, one-third between $-18$ and $-15$ mag, and one-third
  fainter than $-15$ mag -- assuming a cut-off in the LF at
  approximately $-$12 mag.  The observed flattening implies a
  declining contribution from the faintest systems over
  time.\label{fig:faintend}}
\end{figure*}

Such analyses typically rely on large surveys that identify galaxies
at a given epoch and quantify their abundance as a function of mass or
luminosity.  The resulting mass and luminosity functions are commonly
represented with Schechter ($\phi^* (L/L^*)^{\alpha}e^{-(L/L^*)}$) or
double power-law ($\phi^* /((L/L^*)^{-\alpha}+(L/L^*)^{-\beta})$) forms
and reveal how galaxy populations grow over time.  The faint-end slope
$\alpha$ characterizes the relative abundance of faint galaxies and
therefore their contribution to the ultraviolet luminosity density
(\S\ref{sec:sources}).

A critical prerequisite for these determinations is accurate redshift
estimates. In the early universe, redshifts are often derived
photometrically using multiwavelength imaging and spectral features
such as the Lyman and Balmer breaks, the 1.6 $\mu$m feature, and
strong lines such as [O III]$_{4959,5007}$ and
H$\alpha$. Spectroscopic follow-up for a subset of sources is
essential to validate and calibrate these photometric estimates.

In the decade before {\it JWST}, luminosity and mass functions were
pushed from $z\sim4$ in 1999 \citep{Steidel1999} to $z\sim7$ in 2004
\citep{Bouwens2004}, and to $z\sim10$ by 2011–2013
\citep{Bouwens2011,Ellis2013,McLure2013,Oesch2013}. Progress then
plateaued
\cite{Bouwens2015,Finkelstein2015,Bouwens2019,Bowler2020,Bouwens2021,Harikane2022}
until {\it JWST} enabled sensitive observations across 1.6–5.0 $\mu$m,
reopening the high-redshift frontier.

Following the release of deep NIRCam imaging, several groups quickly
extended luminosity-function measurements beyond the {\it HST}
frontier, reporting constraints to $z\sim13$
\citep{Donnan2023,Finkelstein2022} and in some cases even $z\sim17$
\citep{Bouwens2023,Bouwens2023_hudf,Harikane2023}.  One of the early
surprises from {\it JWST} results was the discovery of a substantially
larger population of bright star-forming galaxies at $z>9$—within the
first 450 Myr—than predicted by pre-{\it JWST} models.  Fontana et
al. \cite{Fontana2026} provide a comprehensive recent review.

\subsection{Cosmic Star Formation History}

A useful way to track this evolution is through the cosmic
star-formation rate density derived from the observed UV luminosity
functions. Because of cosmic expansion, these values are typically
expressed in comoving units independent of redshift. The strong UV
output and short lifetimes of massive stars allow a straightforward
conversion from UV luminosity to star-formation rate
\citep{Madau1996,Kennicutt1998}.

\begin{figure*}[t]
\centerline{ \psfig{file=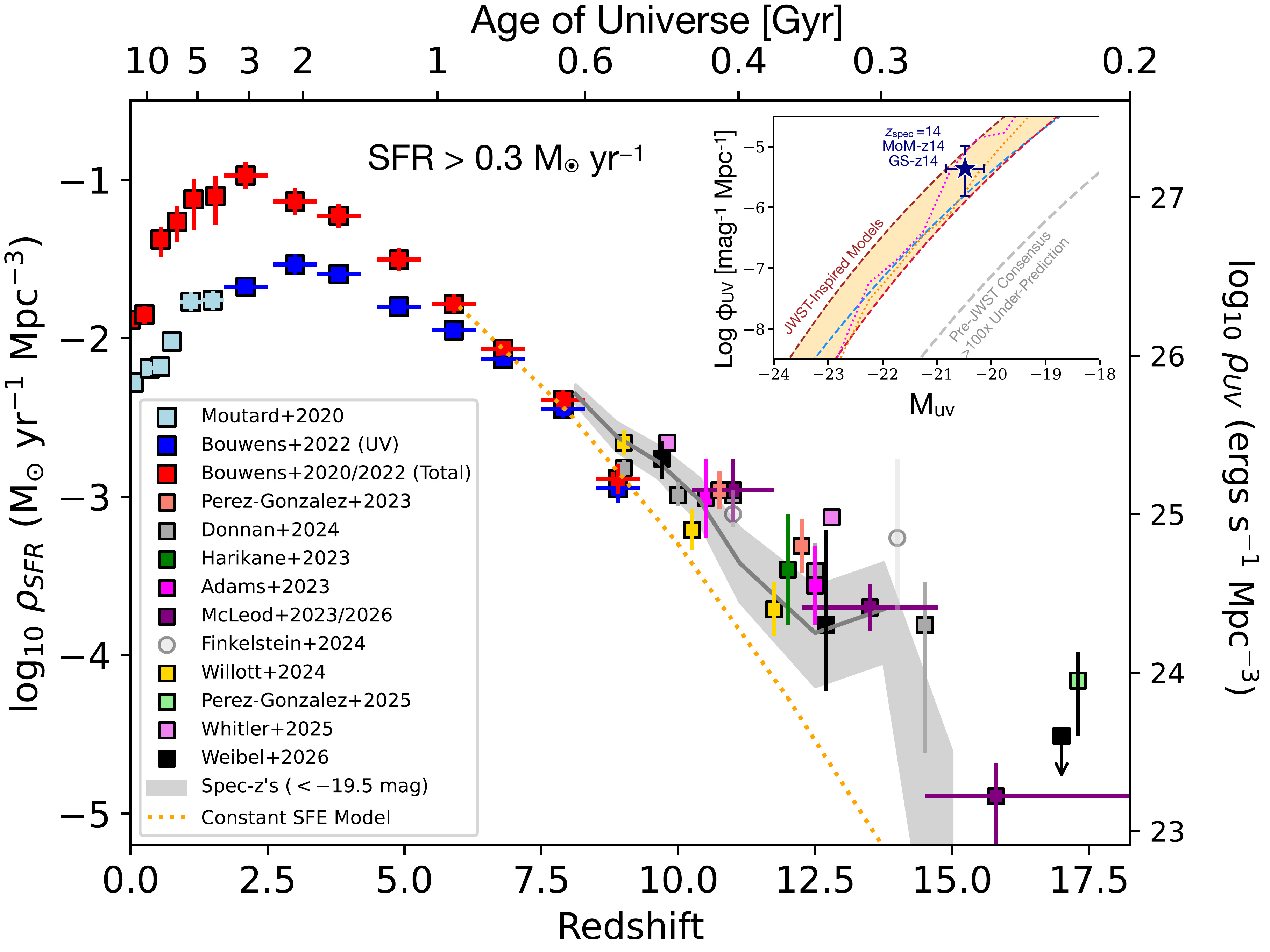,width=0.99\textwidth}}
\vspace{-0.45cm}
\caption{Observational constraints on the cosmic star-formation rate
  density from $z\sim18$ (0.2 Gyr after the Big Bang) to the present
  day ($z\sim0$). Estimates at $z\geq9$ include recent {\it JWST}
  results listed in the legend
  \citep{Donnan2024,Harikane2024,Adams2024,McLeod2024,Finkelstein2024,Willott2024,Weibel2026,PerezGonzalez2025,Whitler2025,McLeod2026}. At
  later times, we show measurements from Bouwens et
  al. \citep{Bouwens2022_hff} spanning $z\sim10$ to $z\sim2$, based on
  rest-UV light with dust corrections, and GALEX plus wide-area
  ground-based surveys from $z\sim2$ to $z\sim0$
  \citep{Wyder2005,Arnouts2005,Moutard2020}. The dotted orange curve
  indicates the expected evolution of star formation in collapsed
  dark-matter halos assuming constant efficiency during the first Gyr
  \citep{Bouwens2015,Mason2015,Mashian2016,Tacchella2018,Harikane2022}. An
  alternative estimate from spectroscopically confirmed samples is
  shown by the gray curve; its closer agreement with the
  constant-efficiency model may suggest that some early candidate
  z$>$10 galaxies lie at lower redshift. As a counterpoint, the
  inset compares the volume density of $z\sim14$ confirmed galaxies
  with fixed-efficiency extrapolations, revealing a $\sim100\times$
  excess (from Naidu et al. \citep{Naidu2026}).\label{fig:sfrd-evol}}
\vspace{-0.3cm}
\end{figure*}

Figure~\ref{fig:sfrd-evol} illustrates several notable measurements of
the cosmic star-formation rate density from {\it JWST} and earlier
studies.  A useful pre-{\it JWST} baseline assumes that a constant
fraction of the gas associated with newly collapsed dark-matter halos
turns into stars.  This produces the dotted orange curve in the
figure; departures from it can signal changing star-formation
efficiency or changes in how observed ultraviolet light maps onto
galaxy abundance. Current estimates at $z>10$ lie well above this
baseline.

Also shown (\textit{shaded grey region}) are estimates derived from
the observed evolution of spectroscopically confirmed bright galaxies
between $z\sim15$ and $z\sim8$ (see also Heintz et
al.\ \citep{Heintz2026}).  Interestingly, these values lie below those
inferred from photometric samples. Because of the interest in
targeting particularly high-redshift galaxies for spectroscopic
confirmation, such selections might nominally be expected to
overestimate the SFR density at $z>12$. The fact that the grey region
instead lies below photometric estimates suggests that the latter may
be overestimated, although sample selection and completeness remain
important uncertainties.

\begin{figure*}[t]
\centerline{
  \psfig{file=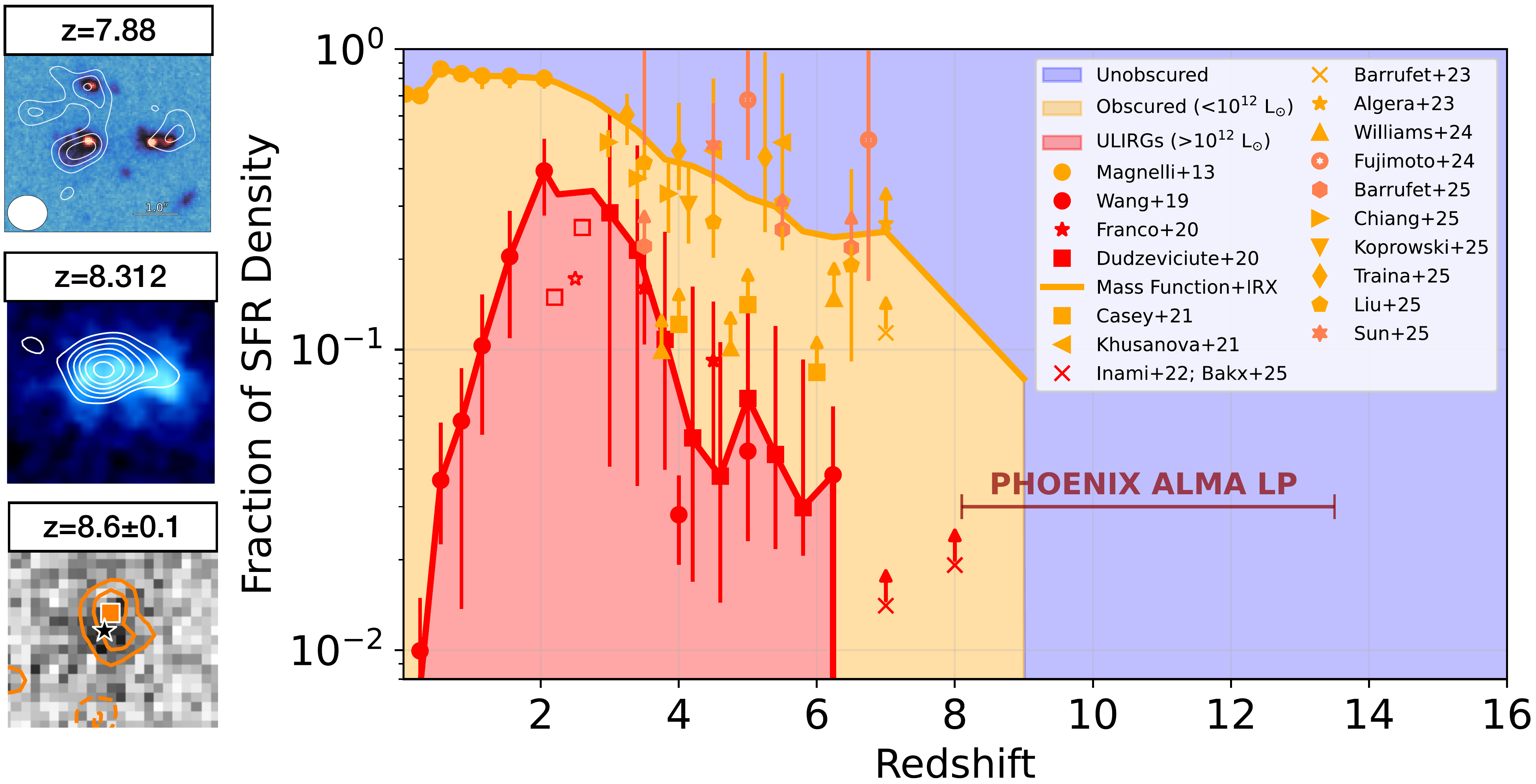
    ,width=0.98\textwidth}}
\vspace{-0.4cm}
\caption{(\textit{Left}) Postage stamps of the three highest-redshift
  dust-detected galaxies currently known, spanning
  spectroscopically-confirmed and photometric redshifts of
  $z\simeq7.9$--8.6
  \citep{Laporte2017,Hashimoto2023,Fudamoto2025,Umehata2025,Tamura2019,vanleeuwen2025}.
  (\textit{Right}) Fraction of the star formation, as a function of
  redshift, that is directly visible in the UV, obscured by dust, or
  dominated by heavily dust-enshrouded systems forming stars at a high
  rate ($>$100 $M_{\odot}$/year: ultra-luminous infrared galaxies:
  ULIRGs
  \citep{Sanders1996,Chary2001,Blain2002,Lonsdale2006,Madau2014})
  \citep{Barrufet2023,Algera2023,Magnelli2013,Khusanova2021,Wang2019Nat,Dudzeviciute2020,Casey2021,Franco2020,Inami2022,Bakx2025,Barrufet2026,Williams2024,Sun2025,Traina2026,Koprowski2025,Liu2026}. The
  fiducial ULIRG contribution presented here follows Magnelli et
  al. \citep{Magnelli2013} at $z=0$–2 and
  Dudzevi{\v{c}}i{\={u}}t{\.{e}} et al. \citep{Dudzeviciute2020} at
  $z=2$–6. The dust-obscured fraction at $z>3$ is estimated by
  applying the Bouwens et al. \citep{Bouwens2020} infrared
  excess–stellar-mass relation (see also
  \citep{Whitaker2017,McLure2018}) to the median stellar-mass
  functions from Weibel et al. \citep{Weibel2024} and Harvey et
  al. \citep{Harvey2025}.  The infrared excess is the ratio of the
  luminosity of galaxies in infrared to the rest-frame UV.  Highly
  dust-obscured galaxies become increasingly rare at high
  redshift. Future observations from the ALMA Large Program PHOENIX
  (PI: Schouws) and related surveys will provide more sensitive
  constraints on the obscured SFR density at $z>8$.
  \label{fig:dustygal}}
\end{figure*}

Spectroscopic confirmation of roughly fifty galaxies in the regime
where the excess was reported establishes that it cannot be explained
solely by contamination from lower-redshift sources.  The confirmed
sources include two bright galaxies at $z>14$---JADES-GS-z14-0
\citep{Carniani2024} and MoM-z14 \citep{Naidu2026}.  The inset to
Figure~\ref{fig:sfrd-evol}, reproduced from Naidu et
al. \citep{Naidu2026}, compares the observed volume density of bright
galaxies at $z\sim14$ with pre-{\it JWST} predictions, showing a
factor-of-$\sim100$ difference.

Several explanations have been proposed for this excess, including
unusually efficient conversion of baryons into stars in the early
universe \citep{Dekel2023,Li2024,Ceverino2024,Somerville2025}, a
top-heavy stellar initial mass function
\citep{Inayoshi2022,Ferrara2023,Schaerer2024,Cueto2024,Chon2024,Trinca2024,Hutter2025,Menon2024,Mauerhofer2025,Cueto2026},
bursty star-formation histories with intermittent quiescent phases
\citep{Mason2023,Sun2023a,Sun2023b,Mirocha2023,Shen2023,Munoz2023,Yung2024,Pallottini2023,Asada2024,Gelli2024sne,Ciesla2024,Nikopoulos2026,Sun2025_bursty,Donnan2025_sfe,Mitsuhashi2026},
contributions from active galactic nuclei
\citep{Inayoshi2022col,Pacucci2022,Volonteri2023,Pacucci2023,Hegde2024},
contributions from nebular continuum emission \citep{Katz2025}, a lack
of dust extinction making galaxies especially $UV$ luminous
\citep{Ziparo2023,Ferrara2023,Ferrara2025,Burgarella2025}, possible
impact of primordial black holes \citep{Liu2025}, and modifications to
the underlying cosmological model
\citep{Menci2022,BoylanKolchin2023,Lovell2023,Shen2024}.

It remains unclear which mechanism is most responsible for the
apparent excess, but it seems likely that more than one contributes.
There are multiple reasons to imagine that bright galaxies like GN-z11
owe their high luminosity to contribution from an AGN
\citep{Maiolino2024,Scholtz2024,Fabian2026,Zhu2026}, and this seems
likely to extend to other extremely compact sources at high redshifts
(Figure~\ref{fig:nitrogen}).  Burstiness must also surely contribute
significantly, given both its observed impact
\citep{Smit2016,Perry2025} and expected greater relevance at very
early times.  Finally, given the youth of the forming galaxies,
nebular continuum emission (Figure~\ref{fig:nebcont}) almost certainly
contributes to the brightness of sources in the rest $UV$
\citep{Katz2025}.

As discussed earlier (\S \ref{sec:sfr}), dust can significantly
obscure starlight when present in sufficient quantities. At very early
times, however, there appears to be a deficit of heavily dust-obscured
galaxies. The highest-redshift galaxy with both a dust-continuum
detection and a spectroscopic redshift remains the $z=8.312$ source
found in 2019 before {\it JWST}'s launch \citep{Tamura2019}.  This is
the only far-infrared detection among 271 spectroscopically confirmed
galaxies at $z>8$ and has a $L_{\rm FIR}>10^{12}$ $L_{\odot}$, with
$>100$ $M_{\odot}\,$yr$^{-1}$ of dust-obscured star formation
\citep{Bakx2025}. Only a handful of additional dust-detected sources
are known in this redshift range, including REBELS-25 at $z=7.306$
\citep{Bouwens2022,Rowland2024} and several QSOs and star-forming
galaxies at $z\sim7.5$--8.6
\citep{Hashimoto2023,Fudamoto2025,Umehata2025,Zavala2026_champs,Barrufet2026,vanleeuwen2025}.

Figure~\ref{fig:dustygal} places these results in a broader context,
showing, as a function of redshift, the fractions of star formation
directly visible in the rest-UV, re-emitted by dust in the
far-infrared, and contributed by ultra-luminous infrared galaxies
(ULIRGs; $>10^{12}$ $L_{\odot}$)
\citep{Magnelli2013,Dudzeviciute2020}. We use galaxy search results
for unobscured star formation from studies of lensing cluster fields
\citep{Bouwens2022_hff} and literature estimates for obscured star
formation
\citep{Barrufet2023,Algera2023,Magnelli2013,Khusanova2021,Wang2019Nat,Dudzeviciute2020,Casey2021,Franco2020,Inami2022,Bakx2025,Barrufet2026,Williams2024,Sun2025,Traina2026,Koprowski2025,Liu2026}. Determining
the prevalence of dust obscuration at these epochs will likely require
dedicated large ALMA programs like REBELS at z$\sim$7
\citep{Bouwens2022} and PHOENIX (PI: Schouws) at even earlier times.

\section{Understanding the Physical Diversity of Galaxies in the Early Universe}

Besides distances to galaxies and measures of their growth, other
properties offer further insight into the nature, evolution, and
environmental impact of early galaxies: their gas content, ubiquitous
ionized-gas emission, chemical enrichment, central black holes, and,
in some cases, a complete absence of ongoing star formation.  These
bear on questions raised in Section 1: how quickly successive
generations of stars enriched early galaxies with metals, their supply
of ionizing radiation, and the emergence and growth of supermassive
black holes.

\begin{figure*}[t]
\centerline{
\psfig{file=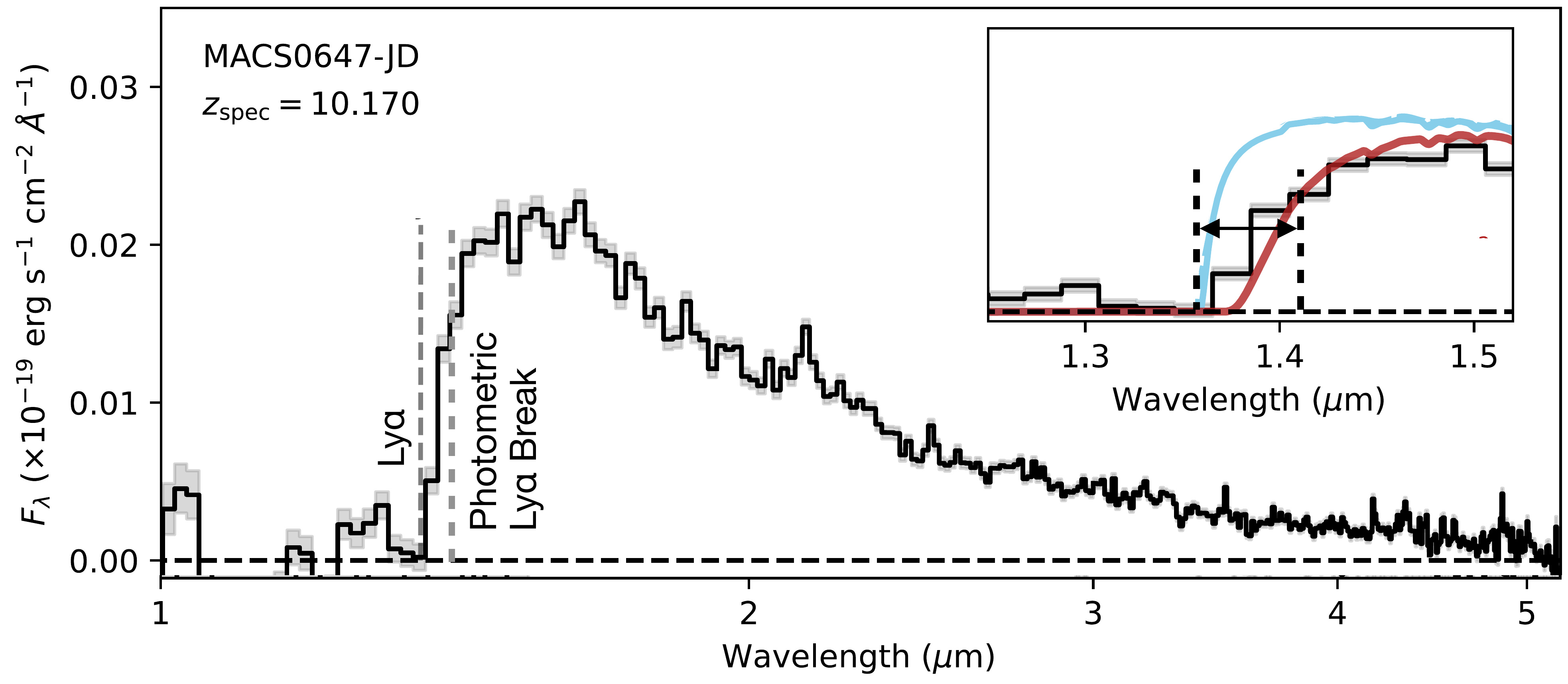,width=0.8\textwidth}}
\vspace{-0.3cm}
\caption{NIRSpec PRISM spectrum of gravitationally lensed galaxy
  MACS0647-JD (z=10.17: \citep{Hsiao2024,Heintz2024_dla}), showing
  strong evidence for a substantial reservoir of neutral hydrogen gas.
  The gas content is inferred from strong Ly$\alpha$ damping of the
  $UV$-continuum light (compare the red line with the light blue line
  in the figure inset).  The vertical dashed line marks the systemic
  redshift, $z=10.17$, obtained from emission lines, whereas the
  broadened break alone would suggest $z\simeq10.6$ (vertical dashed
  line offset to longer wavelengths).  The visible offset between
  these two positions is the signature of absorption by neutral
  hydrogen.\label{fig:dla}}
\end{figure*}

\subsection{High Gas Content and Ionized Gas Emission}

{\it JWST} has provided an unprecedented view of the physical
conditions in early star-forming galaxies, revealing their ionized and
neutral gas content, massive hot stars, and overall interstellar
medium properties. One early clue came from systematic offsets between
redshifts inferred from nebular emission lines and those estimated
from the Ly$\alpha$ break
\citep{Hsiao2024,Heintz2024_dla}. Figure~\ref{fig:dla} shows a
high-S/N example. While the Lyα break suggests $z=10.6\pm0.3$, several
emission lines place the galaxy at $z=10.17$ \citep{Hsiao2024}, about
4\% lower.  Although damping by a neutral intergalactic medium (IGM)
could in principle produce such offsets, even a fully neutral IGM
cannot account for the magnitude of the observed shift. The absorption
is therefore attributed to substantial neutral hydrogen within that
galaxy itself, implying that massive gas reservoirs were already in
place a few hundred million years after the Big Bang
\citep{Heintz2024_dla}. Current studies indicate gas masses that are
5–10$\times$ larger than the stellar masses, including in
JADES-GS-z14-0 at $z=14.18$ \citep{Heintz2025_z14} and other systems
at $z=7$–11 \citep{Heintz2024_dla,Rowland2026_dla}.

\begin{figure*}[t]
\centerline{
\psfig{file=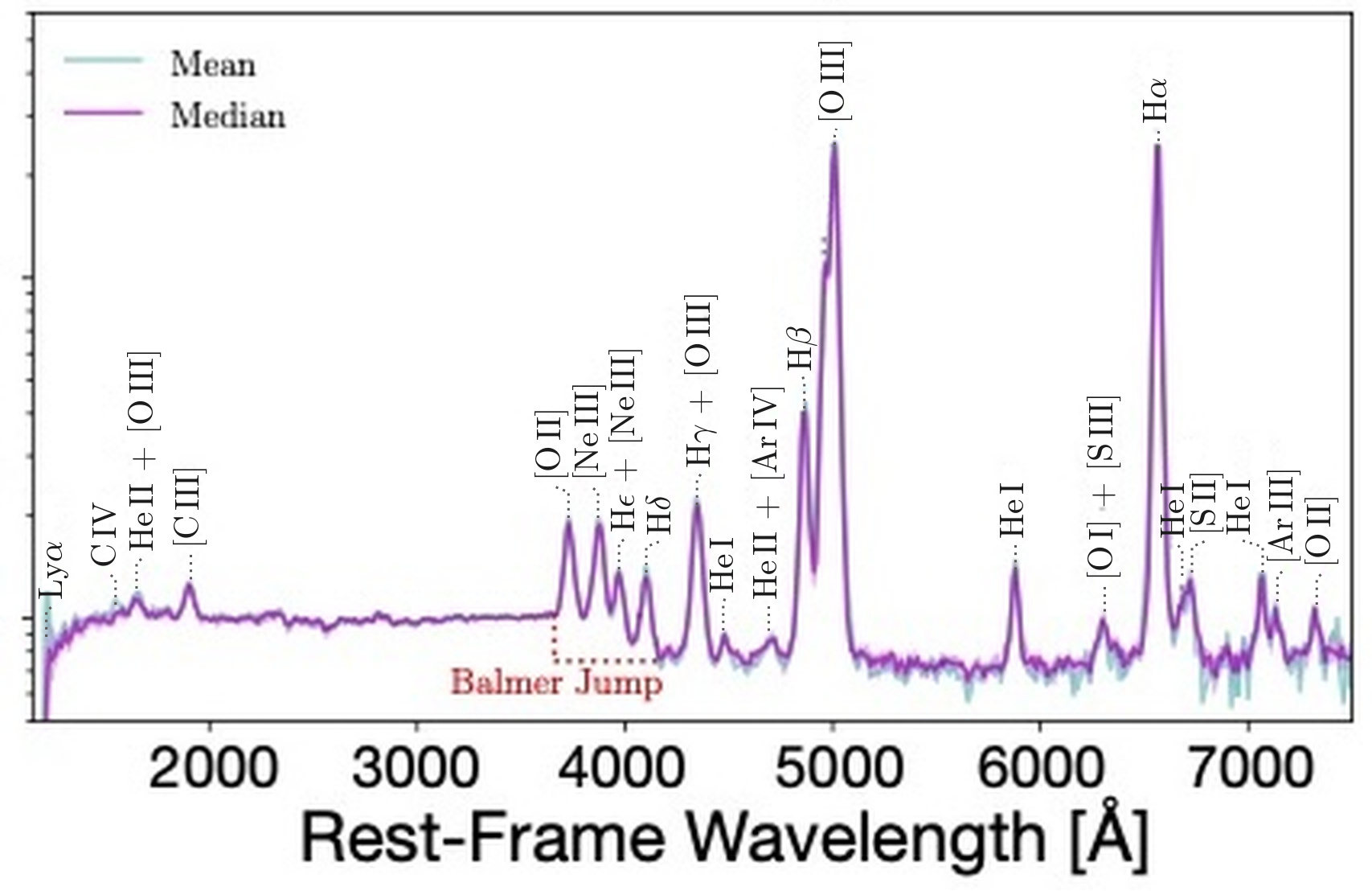,width=0.53\textwidth}}
\vspace{-0.4cm}
\caption{Observed nebular continuum and line emission from young
  star-forming galaxies identified with {\it JWST}: mean and median
  rest-frame spectra stacked from 58 early-universe galaxies, showing
  a strong Balmer jump (the sharp rise in continuum level near
  3646\AA, produced by hot ionized gas surrounding young stars) and
  numerous emission lines as indicated on the figure. Adapted from
  Katz et al. \citep{Katz2025}.  Although such intense nebular and
  line emission is rare in nearby galaxies, it is much more prevalent
  in the early universe.\label{fig:nebcont}}
\end{figure*}

Spectroscopic studies of early star-forming galaxies have shown that
their spectra can be dominated by line and nebular-continuum emission
from ionized gas. Figure~\ref{fig:nebcont} illustrates this emission
using a stack of 58 galaxy spectra with prominent nebular continua.
Strong nebular-continuum emission, including a Balmer jump in the
continuum at $0.36\,\mu\mathrm{m}$, is powered by intense ionizing
radiation from massive, short-lived stars. Although this signature is
uncommon in nearby galaxies, it is much more prevalent in the early
universe, consistent with the idea that young, massive stars in these
systems produce many of the photons responsible for cosmic
reionization (\S\ref{sec:reionization}).

\subsection{Chemical Enrichment\label{sec:metals}}

Ordinary matter in the early universe consisted almost entirely of
hydrogen and helium, with only traces of heavier elements, which
astronomers call ``metals.''  Over cosmic time, metals (e.g., carbon,
oxygen, and nitrogen) are known to be produced in stars through fusion
reactions and eventually returned to the surrounding gas through
stellar winds and supernova explosions.  As a result, we expect a
general monotonic increase in the metal content with cosmic time, and
for metallicity probes to provide us with important information on how
rapidly galaxies converted their gaseous material into stars, while
building the necessary materials for dust and planet formation
\citep{Maiolino2019,Sanders2020,Curti2020}.

\begin{figure*}[t]
\centerline{
  \psfig{file=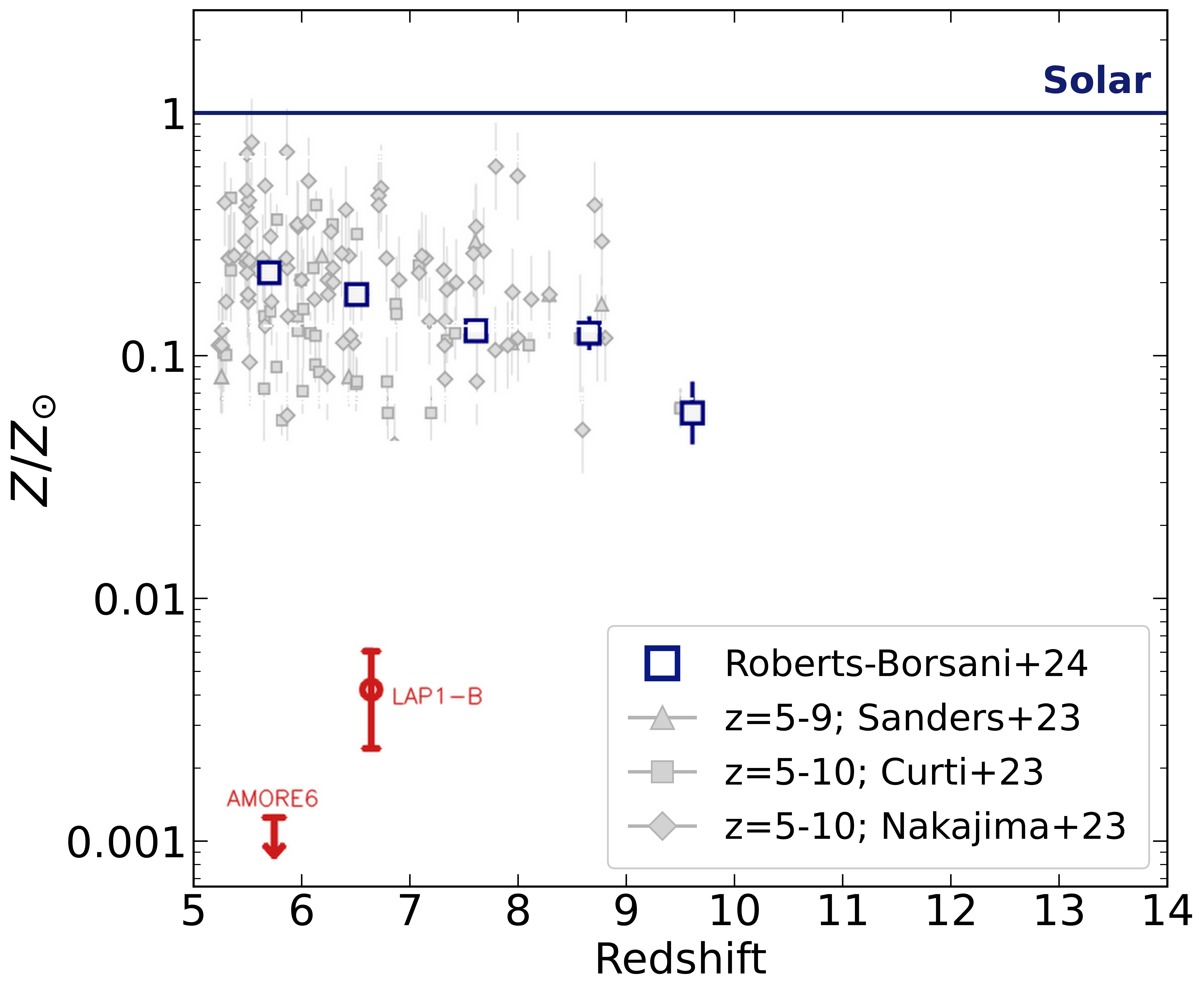,width=0.5\textwidth}}
\vspace{-0.3cm}
\caption{Gas-phase oxygen abundance as a function of redshift.  The
  vertical axis gives the oxygen metallicity, i.e., the gas-phase
  oxygen-to-hydrogen abundance relative to that of the Sun
  ($Z/Z_{\odot}$, logarithmic scale), with the solar value marked by
  the bold horizontal line. Grey points show individual-galaxy
  measurements from the literature, while blue squares show stacked
  measurements in five redshift bins spanning $z=5.5$--9.5, based on
  strong emission-line ratios for a benchmark sample of nearly 500
  spectroscopically confirmed {\it JWST} galaxies
  \citep{RobertsBorsani2024}.  The stacked measurements reveal a
  clear, gradual decline in typical metallicity with increasing
  redshift, consistent with progressively less chemically processed
  gas.  The vertical axis has been extended to
  $\sim$0.001\,$Z_{\odot}$ to additionally show the extremely
  metal-poor sources LAP1-B \citep{Nakajima2026} and AMORE6
  \citep{Morishita2025} (red symbols; AMORE6 is an upper limit), which
  lie more than an order of magnitude below the Roberts-Borsani et
  al. stacked trend.\label{fig:metaltrend}}
\end{figure*}

Before {\it JWST}, metallicity measurements at $z>3$ were limited due
to the need for rest-frame optical spectroscopy.  Select studies used
far-infrared ISM cooling line with ALMA \citep{Jones2020}, but only a
small number of such measurements could be made.  {\it JWST}’s
sensitive NIRSpec coverage from 2.3–5.2 $\mu$m has completely
revolutionized this field, enabling measurements of key rest-frame
optical lines such as [O II]$_{3727,3729}$, H$\beta$, [O
  III]$_{4959,5007}$, and the temperature-sensitive auroral line [O
  III]$_{4363}$
\citep{Curti2023,Sanders2024,Laseter2024,Morishita2024,Sanders2026,Cataldi2025,Rogers2026}. As
a result, metallicities have now been derived for many
hundreds of galaxies at $z=5$–9
\citep{Curti2023,Tang2023,Heintz2023_FMR,Nakajima2023,Fujimoto2023,Curti2024,Chemerynska2024,Morishita2024,RobertsBorsani2024,Sarkar2025}. The
derived metallicities are broadly lower than at later epochs,
consistent with significant inflows from pristine gas.  However,
near-solar metallicities are still seen for the most massive galaxies
at these redshifts \cite{Shapley2025,Rowland2026,Faisst2026,Chen2026a},
suggesting that rapid enrichment is possible for the most massive
systems.  Figure~\ref{fig:metaltrend} places these measurements in a
broader statistical context, showing how the {\it typical} metallicity
of {\it JWST}-confirmed galaxies declines gradually with redshift,
from roughly one-fifth solar at $z\sim5.5$ to $\sim$5\% solar by
$z\sim9.5$.

At $z\geq10$, measurements remain sparse because MIRI spectroscopy is
needed to access [O III]$_{4959,5007}$ and H$\alpha$
\citep{Hsiao2024_MIRI,AlvarezMarquez2025,AlvarezMarquez2026,Zavala2025,Helton2026}. Fortunately,
thanks to on-going spectroscopy of $>10$ luminous $z>10$ galaxies
\citep{AlvarezMarquez2025_jwst,Endsley2025_jwst}, the numbers are
continuing to increase.  Where determinations are possible,
metallicities are typically $\lesssim$10\% of the solar value
($\lesssim$0.1\,$Z_{\odot}$), consistent with progressive chemical enrichment from
ongoing star formation and supernova activity.

\begin{figure*}[t]
\centerline{
  \psfig{file=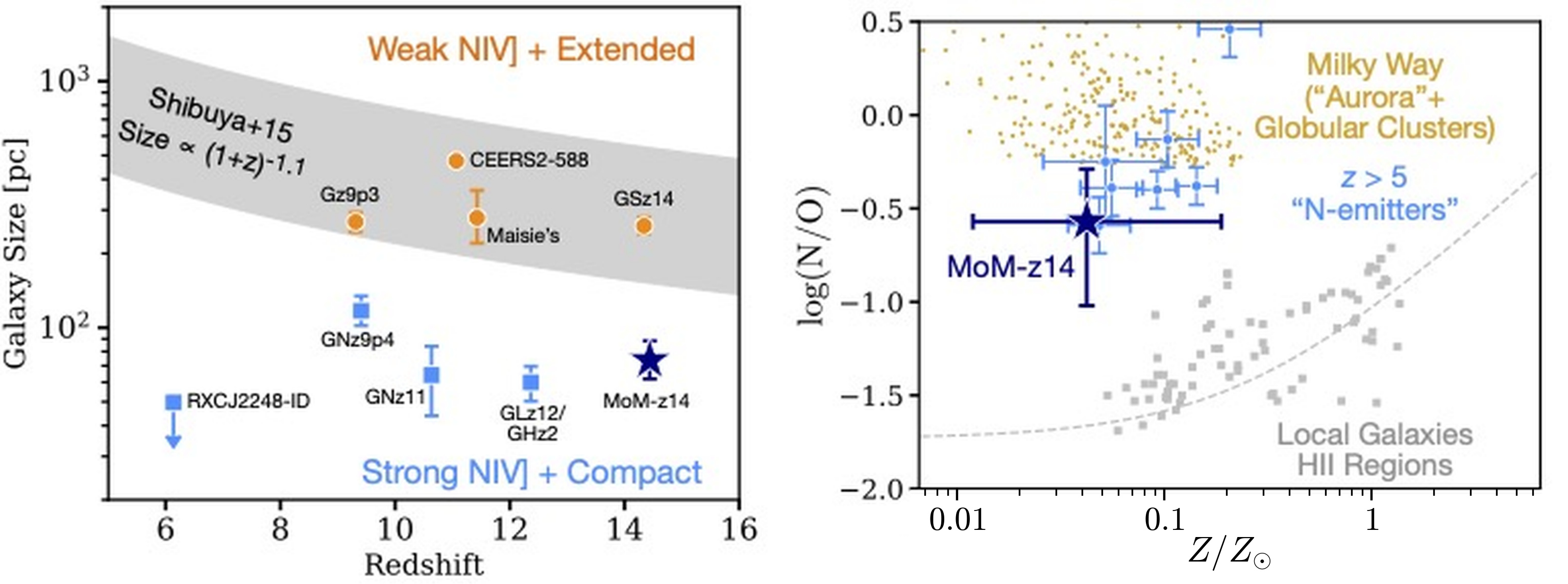,width=0.99\textwidth}}
\vspace{-0.5cm}
\caption{(\textit{Left panel}) Inferred spatial radii of notable
  $z>5.5$ galaxies as a function of spectroscopic redshift. Systems
  with strong NIV] emission are shown in blue, while those with weaker
NIV] emission are shown in orange. Galaxies with weak NIV] emission
    follow expected lower-redshift size trends, whereas strong
    emitters are typically much more compact
    \citep{Harikane2025,Naidu2026}. Both panels, adapted from Naidu et
    al. \citep{Naidu2026}, highlight the location of MoM-z14 at
    $z=14.44$. (\textit{Right panel}) Observed [N/O] abundances of
    NIV]-bright galaxies as a function of oxygen metallicity relative
      to solar ($Z/Z_{\odot}$), compared with local galaxies and H II
      regions (grey points). The enhanced [N/O] ratios resemble those
      of Milky Way globular clusters, suggesting a similar stellar
      formation pathway.  \label{fig:nitrogen}}
\end{figure*}

One surprising finding is the apparently high nitrogen abundance in some
luminous $z>8$ galaxies (Figure~\ref{fig:nitrogen}). This was first
noted in GN-z11 from its strong N III]$_{1747,1749}$ and N
  IV]$_{1486}$ emission lines
    \citep{Bunker2023_gnz11,Cameron2023}. Similar nitrogen enhancement
    has since been found in a subset of galaxies spanning
    $z\simeq8.6$--14.4
    \citep{Morishita2026_bubble,MarquesChaves2024,Schaerer2024,Curti2025,Castellano2024,Naidu2026,Morel2025,Cameron2026,Watanabe2026}.
    These systems often host intense starbursts
    \citep{Topping2024,Topping2025,Martinez2025,Berg2026} and are
    unusually compact \citep{Harikane2025,Naidu2026}. Similar chemical
    patterns occur in Milky Way globular clusters, suggesting that
    dense star-forming environments may favor efficient nitrogen
    production \citep{Gratton2004,Bastian2018}.

Given the increasingly pristine gas expected in the highest-redshift,
lowest-mass regions of the universe, there has been substantial
interest in identifying extremely low-metallicity galaxies containing
Population III stars---the hypothetical first generation formed from
pristine hydrogen and helium.  No true Population III galaxy has yet
been confirmed, but here we highlight two examples that come close
(see also
\citep{Hsiao2025_lowmetal,Fujimoto2025_pop3,Willott2025,Cullen2025}).
Both examples lie well below the trend in Figure~\ref{fig:metaltrend},
with oxygen abundances of about 0.4\% and less than 0.12\% of the
solar value, respectively. These values are more than an order of
magnitude below that of a typical galaxy at a similar redshift,
underscoring how unusual these systems are.

The first is a highly magnified star cluster LAP1-B at $z=6.625$.
Earlier observations gave an upper limit on its oxygen abundance
\citep{Vanzella2023}; deeper, dedicated NIRSpec spectroscopy now gives
0.004$\pm$0.002 $Z_{\odot}$ \citep{Nakajima2026}
(Figure~\ref{fig:metaltrend}).  This highly magnified source was
identified from deep observations of Hubble Frontier Fields cluster
MACS0416.  It reveals a very small star-forming region and is a
potential proto-globular cluster \citep{Vanzella2020,Vanzella2021}.
The second example is AMORE6 at $z=5.725$, identified by Morishita et
al. \citep{Morishita2025} behind the lensing cluster Abell 2744.  Both
systems are notable for showing line emission in H$\beta$ or
H$\alpha$, but little detectable [O III]$_{4959,5007}$ emission.  This
indicates that, while the sources produce a substantial number of
high-energy photons capable of powering Balmer line emission, their
oxygen abundance is too low to produce strong [O III]$_{4959,5007}$
emission.

The fact that both of these discoveries occur at $z\sim6$--7 is partly
due to the wavelength coverage and sensitivity of {\it JWST}. At these
redshifts, {\it JWST}/NIRSpec can observe both H$\alpha$ and the [O
  III]$_{4959,5007}$ doublet with high sensitivity.  The existence of
such systems at $z\sim6$--7 need not be surprising, because chemical
enrichment depends not only on cosmic time but also on a galaxy's
mass, star-formation history, environment, and inflow of relatively
unprocessed gas. A low-mass galaxy that has formed few stars, or that
is continually replenished with nearly pristine gas, can therefore
remain chemically primitive long after more massive galaxies nearby
have become enriched.

\subsection{Supermassive Black Holes}

Finding rapidly growing black holes in the first billion years is
surprising because there is little time for an initial ``seed'' black
hole to form and grow to millions of solar masses \citep{Fan2023}.
Their masses are usually inferred from the speed of gas close to the
black hole, measured through the width of bright hydrogen lines,
together with the luminosity of the central source.  Broad permitted
lines trace fast-moving gas near the black hole, whereas narrower
forbidden lines usually arise farther away.  {\it JWST} has not only
confirmed billion-solar-mass black holes in the early universe
\citep{Marshall2023,Yue2024_EIGER}, but has also extended measurements to much lower
masses and more typical systems \citep{Onoue2026}, while revealing a
substantial population of dust-obscured QSOs out to $z\sim8.4$ with
MIRI \citep{Lyu2024,Bulichi2026}.

In addition, {\it JWST} has revealed a seemingly ubiquitous population
of what appear to be moderately massive black holes in galaxies
\citep{Labbe2023,Kocevski2023,Harikane2023LRD,Matthee2024,Labbe2025}. These
sources, now known as “Little Red Dots” (LRDs) \citep{Matthee2024},
are largely unresolved, with red spectral slopes in the rest-frame
optical but blue in the rest-frame UV \citep{Kocevski2025}.  Their red
colors and extreme hydrogen line ratios were initially attributed to
dust \citep{Polletta2024,Hainline2024}, but their faintness in the
mid- to far-IR \citep{Yue2024,Setton2025,Xiao2025,Casey2025,Akins2025}
challenged this explanation. Combined with strong Balmer absorption
features, the observations instead point to extremely dense gas
surrounding moderate-mass accreting black holes
\citep{Inayoshi2025,Naidu2026_BHS,deGraaff2025}. Such dense gas can
naturally explain the red optical colors, unusual hydrogen line
ratios, and absorption features.  Using the wide areas already imaged
with {\it JWST}/NIRCam, hundreds of LRDs have now been identified,
mostly at $z\sim4$–7
\citep{Kokorev2024,Kocevski2025,Akins2025}. Remarkably, over 120 have
NIRSpec spectroscopy, providing key constraints on their nature
\citep{deGraaff2025}. Revised analyses -- without invoking heavy dust
obscuration -- now suggest substantially lower black hole masses than
initially estimated \citep{Greene2026,Rusakov2026}.

\begin{figure*}[t]
\centerline{
  \psfig{file=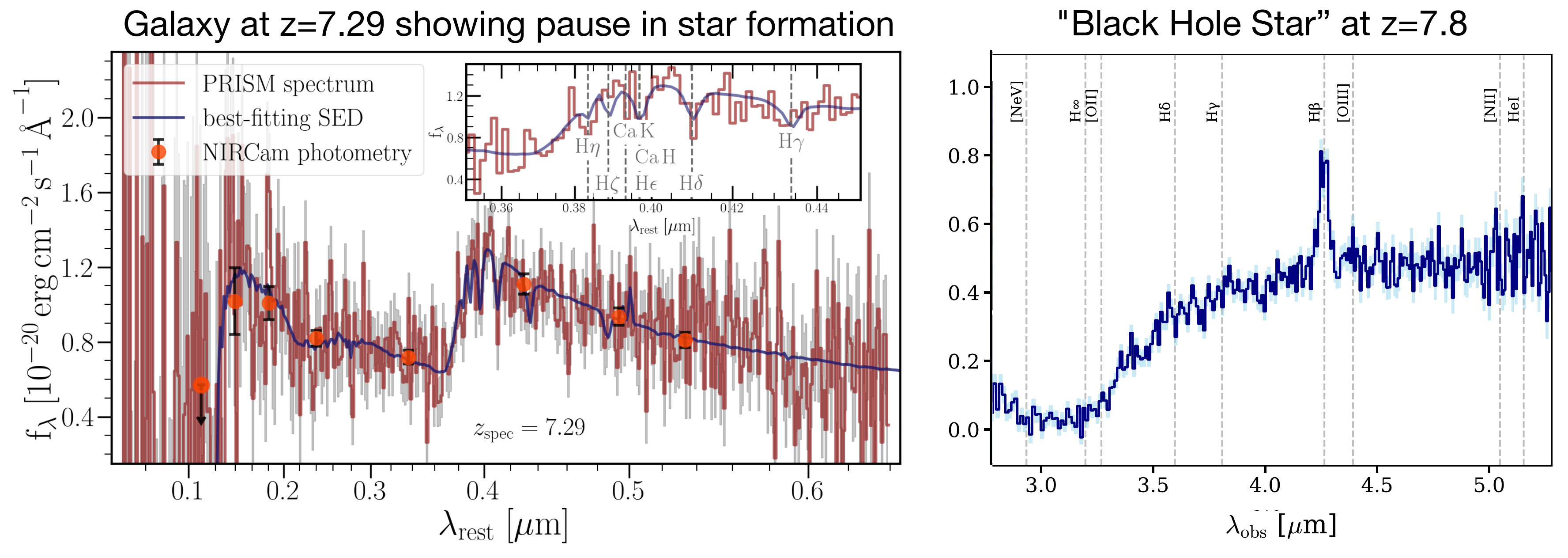,width=0.92\textwidth}}
\vspace{-0.5cm}
\caption{Early-universe discoveries include a $z=7.29$ system with
  little star formation over the past 100 Myr (\textit{left}) and a
  $z=7.8$ candidate ``black-hole star,” representative of the “little
  red dot” population newly identified with {\it JWST}. The left panel
  (from Weibel et al. \citep{Weibel2025_z7p3}) shows NIRCam
  photometry and a NIRSpec PRISM spectrum of the quiescent galaxy,
  with an inset highlighting Calcium H\&K and Balmer absorption lines
  typical of post-starburst systems. The right panel (from Naidu et
  al. \citep{Naidu2026_BHS}) shows a {\it JWST}/NIRSpec spectrum of
  what may be a SMBH (or dense young star cluster) embedded in
  extremely dense neutral hydrogen gas
  \citep{Naidu2026_BHS,deGraaff2025,Inayoshi2025}.\label{fig:diversity}}
\end{figure*}

\subsection{Quiescent Galaxies}

Finally, a number of galaxies have been identified in the early
universe where star formation appears to have stopped for $\gtrsim$50
million years. Such sources are known as quiescent galaxies, and given
rapid gas accretion and cooling at early times, they are expected to
be rare, making up <1\% of distant galaxies
\citep{Carnall2023,Strait2023,Looser2024,Weibel2025_z7p3,Baker2025,CoveloPaz2026,Endsley2025}.

Nonetheless, the discovery of several quiescent galaxies at $z>7$—only
$\sim$750 Myr after the Big Bang—is striking
\citep{Looser2024,Weibel2025_z7p3,Baker2025,Harikane2026}, including
one at $z=11.04$ \citep{Harikane2026}. Their apparent abundance
exceeds predictions from conventional gas accretion and star-formation
models \citep{Carnall2023,Weibel2025_z7p3,Baker2025_sample},
suggesting that gas supply is not the primary limitation. Instead,
strong feedback may temporarily halt star formation \citep{Gelli2024}
until fresh cold gas is accreted.  Here feedback means radiation,
energy, and outflowing material from stars, supernovae, or accreting
black holes that heat or expel gas and thereby regulate later star
formation. Such feedback likely arises from AGN
in massive systems and supernova activity in lower-mass galaxies
\citep{Xie2024,Gelli2024sne}. Deeper spectroscopic and photometric
surveys \citep{Baker2025_sample,CoveloPaz2025jwst} will be crucial for
constraining the prevalence and physical drivers of these quiescent
galaxies at high redshift.

\begin{figure*}
\centerline{
  \psfig{file=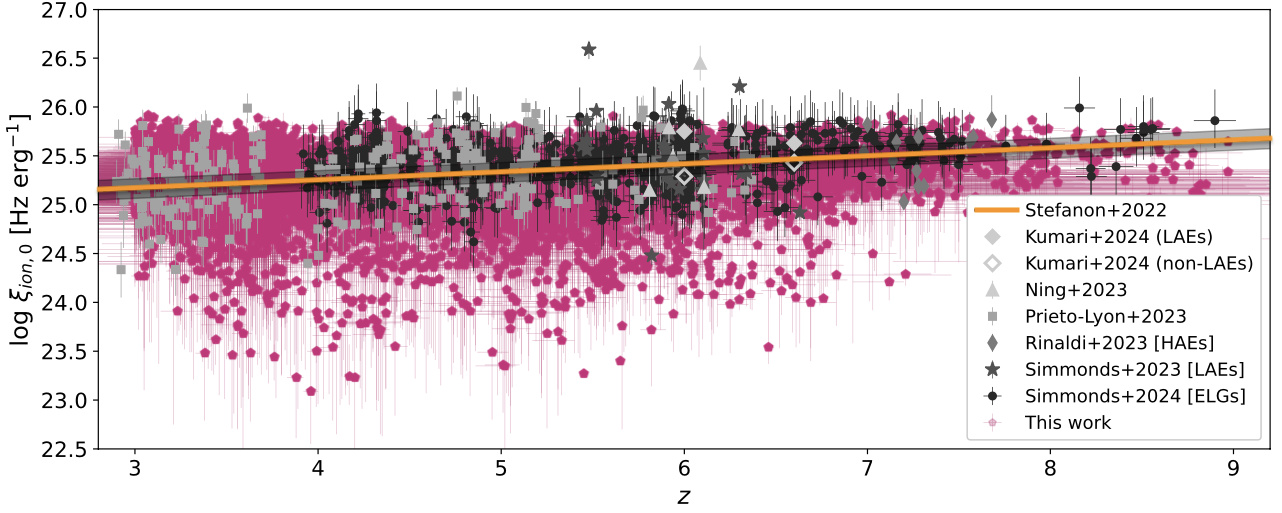,width=0.98\textwidth}}
\vspace{-0.65cm}
\caption{Measurements of $\xi_{\rm ion}$—the number of ionizing
  photons produced per unit rest-frame UV luminosity in star-forming
  galaxies.  Shown are pre-{\it JWST} estimates (\textit{orange solid
    line}) from ground-based spectroscopy and {\it Spitzer}/IRAC as
  compiled by Stefanon et al.\ \citep{Stefanon2022}
  including results from many studies
  \citep{Fumagalli2012,Sobral2014,Bouwens2016,Smit2016,Marmol2016,Rasappu2016,Matthee2017,Harikane2018,Reddy2018,Faisst2019,Nanayakkara2020,Atek2022}. {\it
    JWST}-based measurements include Kumari et al. \citep{Kumari2024},
  Ning et al. \citep{Ning2023}, Prieto-Lyon et
  al. \citep{PrietoLyon2023}, Rinaldi et al. \citep{Rinaldi2024},
  Simmonds et al. \citep{Simmonds2023}, Simmonds et
  al. \citep{Simmonds2024a}, and Simmonds et
  al. \citep{Simmonds2024}.\label{fig:xiion}}
\end{figure*}

\section{Cosmic Reionization\label{sec:reionization}}

Following recombination, which occurred 3.8$\times$10$^5$ years after
the Big Bang, the universe transitioned to a largely neutral state,
with only $\sim$0.01\% of the available hydrogen remaining in an
ionized state \citep{Peebles1968}. Today, however, intergalactic
hydrogen is almost completely ionized, so high-energy radiation—likely
from massive stars and accreting supermassive black holes—must at some
point have reionized it \citep{Loeb2001}.  This period is referred to
as the epoch of reionization.

Observations indicate that the ionization of the ambient hydrogen was
largely complete $\sim$1 billion years after the Big Bang
\citep{Fan2006_review,Fan2023,Dayal2018}.  Key constraints include the
impact of Thomson scattering on the cosmic microwave background power
spectrum \citep{WMAP2013,Planck2016,Planck2020,Planck2020_A1}, the
Gunn--Peterson trough, a broad region of almost complete absorption
blueward of Ly$\alpha$, observed in the spectra of bright quasars at
z$\geq$6 \citep{Becker2001,Fan2001,Fan2002,Fan2006}, and the effect of
ambient neutral hydrogen in creating Ly$\alpha$ damping wings and
attenuating Ly$\alpha$ photons \citep{Robertson2013,Fan2023}.  Prior
to {\it JWST}, the prevailing view was that high-mass stars in
star-forming galaxies provided the bulk of the ionizing photons to
reionize the universe \citep{Robertson2022}.

{\it JWST} has significantly improved our ability to characterize both
the sources of ionizing photons and the neutrality of hydrogen during
reionization. It detects galaxies $\sim$6$\times$ fainter than {\it
  HST} could, and has provided large spectroscopic samples at $z>6$,
extending to $z=14.44$ (Figure~\ref{fig:jwst-z-machine}).

\subsection{Sources of Ionizing Photons\label{sec:sources}}

\begin{figure*}
\centerline{
  \psfig{file=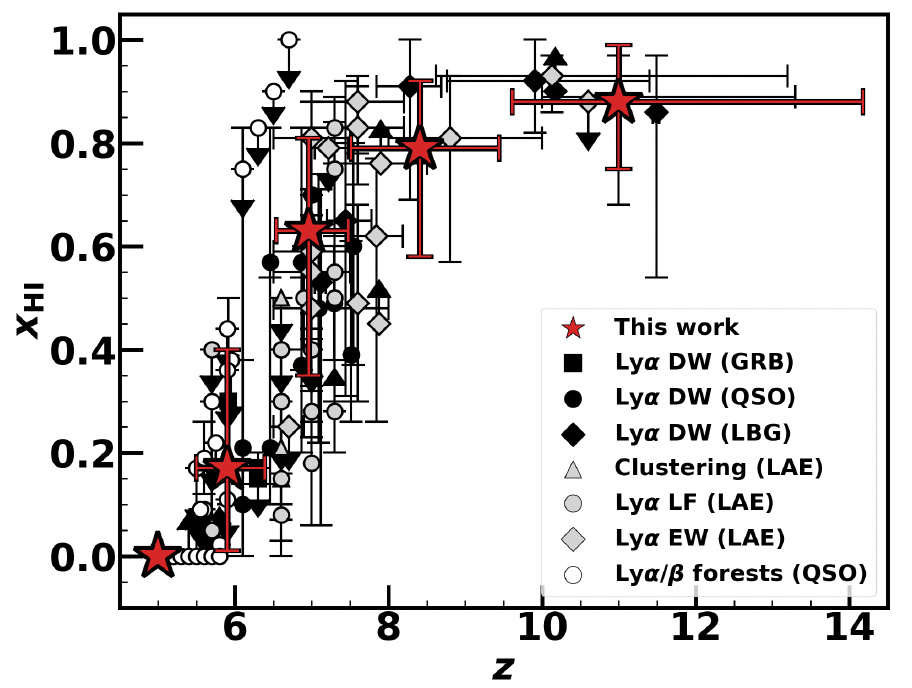,width=0.6\textwidth}}
\vspace{-0.3cm}
\caption{Measurements of the neutral hydrogen fraction in the
  early-universe intergalactic medium, used to model reionization
  driven by star-forming galaxies. Shown are constraints from
  Ly$\alpha$ damping wings seen in the $UV$-continuum spectra of
  gamma-ray bursts \citep{Totani2006,Totani2014,Fausey2025}, quasars
  \citep{Davies2018,Durovcikova2020,Durovcikova2024,Wang2020,Yang2020,Greig2022},
  and Lyman-break galaxies \citep{CurtisLake2023,Hsiao2024,Umeda2024},
  and from the prevalence of Ly$\alpha$ emission
  \citep{Ouchi2010,Konno2014,Konno2018,Zheng2017,Inoue2018,Goto2021,Morales2021,Ning2022},
  the clustering of Ly$\alpha$-emitting galaxies
  \citep{Ouchi2010,Ouchi2018,Sobacchi2015}, and Ly$\alpha$ equivalent
  width distribution in star-forming galaxies
  \citep{Mason2018,Mason2019,Hoag2019,Jung2020,Whitler2020,Bolan2022,Morishita2023,Bruton2023,Jones2024,Nakane2024,Tang2024}
  including a determination (red stars) based on 586 galaxies at
  $z=4.5$-14.2 \citep{Kageura2025}.  Both probes are highly sensitive
  to modest neutral-hydrogen columns in the IGM. Also shown are
  constraints from the fraction of spectral regions completely
  absorbed by neutral hydrogen
  \citep{McGreer2015,Zhu2022,Zhu2024,Jin2023,Spina2024}.  This figure
  is taken from Kageura \citep{Kageura2025}.
  \label{fig:reionization}}
\end{figure*}

It is standard
\citep{Madau1999,Robertson2013,Bouwens2015_reion,Bouwens2016,Robertson2022}
to estimate the ionizing output of star-forming galaxies as the
product of three terms: (1) $\rho_{\rm UV}$, the UV luminosity density;
(2) $\xi_{\rm ion}$, the ionizing photon production efficiency; and (3)
$f_{\rm esc,\mathrm{LyC}}$, the fraction of ionizing photons that escape
into the IGM. In essence, $\rho_{UV}$ measures the total UV brightness
per unit comoving volume.

Because no survey can detect arbitrarily faint galaxies, $\rho_{\rm
  UV}$ must be extrapolated below the observational limit.  A key
parameter in this extrapolation is the faint-end slope $\alpha$,
defined through the $UV$ luminosity function as $\propto
L^{\alpha}$. For $\alpha=-2$, each logarithmic interval in luminosity
contributes equally, as the higher abundance of faint galaxies offsets
their lower luminosities. Shallower slopes shift the balance to
brighter systems, while for steeper slopes ($\alpha<-2$) numerous
unseen faint galaxies supply a large fraction of the total ultraviolet
light \citep{Bouwens2016_review}.

{\it HST} data already provided meaningful constraints on $\rho_{UV}$
out to $z\sim8.5$ \citep{Stark2016,Robertson2022}, aided by
measurements of $\alpha$. Figure~\ref{fig:faintend} summarizes both
pre-{\it JWST} \citep{Atek2018,Bouwens2022} and new {\it JWST}
constraints
\citep{Leung2023,Chemerynska2026,Harikane2024,Whitler2025,Atek2025_glimpse,Atek2026}. Much
of this progress relied on gravitational lensing by massive clusters,
enabling studies of galaxies $\sim50\times$ fainter than otherwise
accessible
\citep{Alavi2014,Livermore2017,Bouwens2017,Atek2018,Ishigaki2018,Bhatawdekar2019,Bouwens2022,Chemerynska2026,Atek2026}.

While {\it JWST} has improved earlier probes of the z$<$8.5 universe
provided by {\it HST}, its greatest impact has been on the $UV$
luminosity density at $z>8.5$ (Figure~\ref{fig:sfrd-evol}). Despite
extending measurements to $z\sim12$, there is little evidence for
significant evolution in the faint-end slope $\alpha$ between
$z\sim12$ and $z\sim8$ \citep{Leung2023,Whitler2025,Chemerynska2026}
(Figure~\ref{fig:faintend}). Although some studies suggested slower
evolution of the $UV$ luminosity density from $z\sim12$ to $z\sim9$
\citep{Donnan2024,Finkelstein2024}, spectroscopic redshift
distributions (\textit{grey shaded region}) instead suggest a smoother
buildup over time.

The ionizing photon production rate can be estimated from hydrogen
recombination lines, particularly Balmer transitions
\citep{Bouwens2016}. Before {\it JWST}, constraints on $\xi_{\rm ion}$
were largely limited to $z\lesssim2$ \citep{Matthee2017,Shivaei2018},
with tentative estimates at $z\sim4$–5 from {\it Spitzer}/IRAC flux
measurements \citep{Bouwens2016,Rasappu2016,Lam2019,Faisst2019} and rest-$UV$ spectroscopy
\citep{Stark2015,Stark2017,Hutchison2019}. {\it JWST}’s spectroscopic
sensitivity beyond 2.5 $\mu$m now enables measurements of $\xi_{\rm
  ion}$ for hundreds of galaxies at $z>5$
\citep{PrietoLyon2023,Simmonds2024a,Simmonds2024,Papovich2026,Saxena2024_xion}. These
studies
\citep{PrietoLyon2023,Ning2023,Simmonds2023,Simmonds2024a,Rinaldi2024,Simmonds2024,Kumari2024,Saxena2024_xion,Endsley2024,Pahl2025,Atek2023}
find $\xi_{\rm ion}$ values at $z>5$ that are typically
$\sim2$–$3\times$ higher than at $z\sim2$–4
\citep{Matthee2017,Shivaei2018}, consistent with younger, more massive
stellar populations at early times (Figure~\ref{fig:xiion}).

\begin{figure*}
\centerline{
\psfig{file=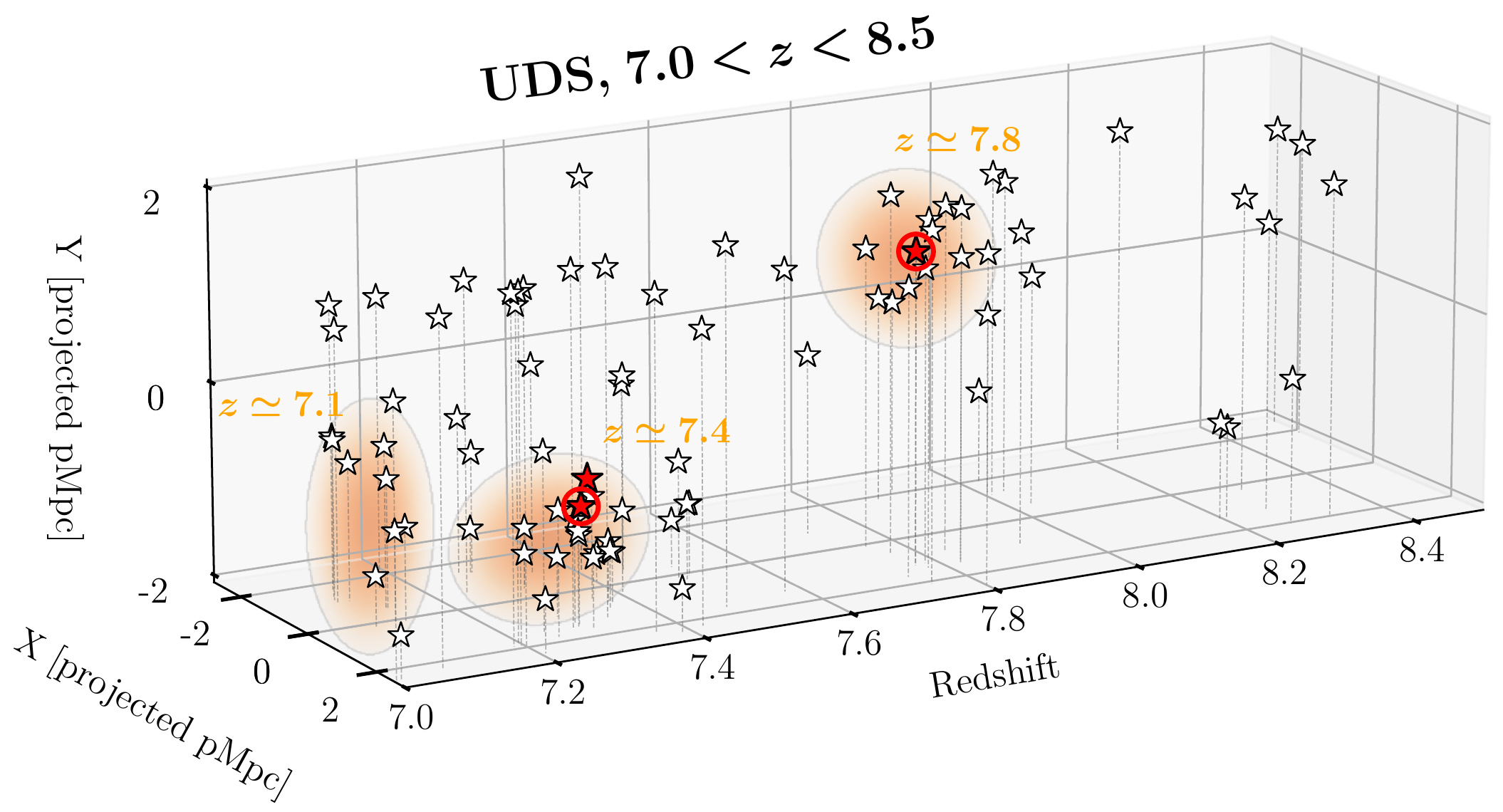,width=0.9\textwidth}}
\vspace{-0.5cm}
\caption{Emerging evidence for ionized bubbles in the early universe inferred from recent analyses \citep{Tang2023,Tang2024,Chen2026_bubbles,Lu2025,Nikolic2025,Martin2026} based on the prevalence of Ly$\alpha$ emission in the spatial distribution of galaxies.  Adapted from Chen et al. \citep{Chen2026_bubbles}, the figure shows the PRIMER UDS field spanning redshifts 8.5 to 7.0. Open stars mark spectroscopically confirmed star-forming galaxies, while red stars denote systems with detected Ly$\alpha$ emission. Because a high neutral fraction suppresses Ly$\alpha$ transmission, these emitters likely reside within ionized regions. Notably, Ly$\alpha$ sources cluster in overdense environments, suggesting that the densest regions ionize first.
  \label{fig:bubbles}}
\end{figure*}

Finally, the ionizing photons which set $f_{\rm esc}$ have energies
$>$13.6 eV and rest-frame wavelengths $<$912$\,$\AA.  For galaxies in
the reionization era, direct measurements of $f_{\rm esc}$ are
infeasible due to line-of-sight attenuation, so studies often rely on
lower-redshift analogues ($>2$ Gyr after the Big Bang)
\citep{Grazian2017,Jaskot2025}. Moreover, $f_{\rm esc}$ depends
sensitively on the small-scale distribution of neutral gas and can
vary with gas inflows, outflows, and small-scale ISM structure
\citep{Wise2014,Trebitsch2017,Barrow2020,Gnedin2022}.

Empirically constraining $f_{\rm esc}$ has been difficult because the
detection of ionizing radiation from sources is rare and generally
requires space-based observations to improve robustness.  At
$z\sim2$–3, where direct detection is possible, many early claims were
later shown to be contaminated by foreground sources
\citep{Vanzella2010,Siana2015,Mostardi2015,Pahl2021}. Improved
analyses used {\it HST}-resolution imaging to reduce confusion, and
stacked deep spectra near 912\AA\ to detect faint Lyman Continuum
flux. The most robust studies infer escape fractions of $\sim6$–10\%
at $z\sim3$ \citep{Pahl2021}.

An effective strategy for probing $f_{\rm esc}$ has been to target
galaxies that are bright in Ly$\alpha$ (the $n$$=$2$\rightarrow$1
transition), since Ly$\alpha$ radiative transfer is governed by the
same neutral gas that regulates the escape of ionizing photons.
Several studies have found a correlation between Ly$\alpha$ and
ionizing escape fractions \citep{Steidel2018,Pahl2021}. The strong
evolution in Ly$\alpha$ escape from z$\sim$6 to z$\sim$0
\citep{Hayes2011} suggests similar evolution in $f_{\rm esc}$.
Systems whose Ly$\alpha$ line lies near the systemic redshift, or that
show closely spaced blue and red Ly$\alpha$ peaks, are especially
indicative of high escape fractions \citep{Naidu2022_EscapeFraction},
though few such systems \citep{Torralba2024} have thus far been
characterized with {\it JWST}.

{\it JWST}-based studies also compare emission-line ratios to
lower-redshift analogues with direct $f_{\rm esc}$ measurements
\citep{Reddy2016,Steidel2018,Pahl2021,Jaskot2024a,Jaskot2024b},
yielding typical estimates of $\sim10\%$
\citep{Saxena2024_xion,Mascia2024,Gazagnes2025}. Complementary
approaches correlate the apparent IGM neutrality in QSO spectra with
nearby galaxies; while uncertain, these analyses likewise suggest
escape fractions of order $\sim10\%$
\citep{Kakiichi2018,Kakiichi2025,Kashino2026}.

\subsection{Evolving Neutrality of the Ambient Hydrogen in First Gyr}

Before {\it JWST}, several strong constraints already existed on the
evolving neutral hydrogen fraction of the universe
(Figure~\ref{fig:reionization}), primarily from the cosmic microwave
background \citep{WMAP2013,Planck2016,Planck2020} and the observed
Gunn--Peterson trough caused by intervening neutral hydrogen
\citep{Becker2001,Fan2001,Fan2002,Fan2006}.  Additional constraints on
the evolving neutral fraction arise from Ly$\alpha$ damping wings
observed in bright sources like QSOs
\citep{Greig2017,Greig2019,Greig2022,Banados2018,Davies2018,Durovcikova2020,Durovcikova2024,Wang2020,Yang2020}
and from the visibility of Ly$\alpha$ emission from galaxies
\citep{Mason2018,Mason2019,Hoag2019,Whitler2020,Bolan2022,Bruton2023}.
These constraints can now be extended to much larger samples thanks to
the sensitivity and efficiency of {\it JWST} observations
\citep{CurtisLake2023,Hsiao2024,Umeda2024,Jones2024,Nakane2024,Tang2024}.
  
Central to many of these reionization probes is the visibility of
Ly$\alpha$ photons, or continuum light blueward or immediately redward
of Ly$\alpha$, because these signals are extremely sensitive to
neutral hydrogen. Even when only a small fraction of the hydrogen is
neutral, a Ly$\alpha$ photon may be repeatedly absorbed and re-emitted
as it passes through the gas, scattering it out of our line of
sight. The line is therefore much easier to detect when a galaxy lies
within a sufficiently large ionized region, or ``bubble.''

\begin{figure*}[t]
\centerline{
  \psfig{file=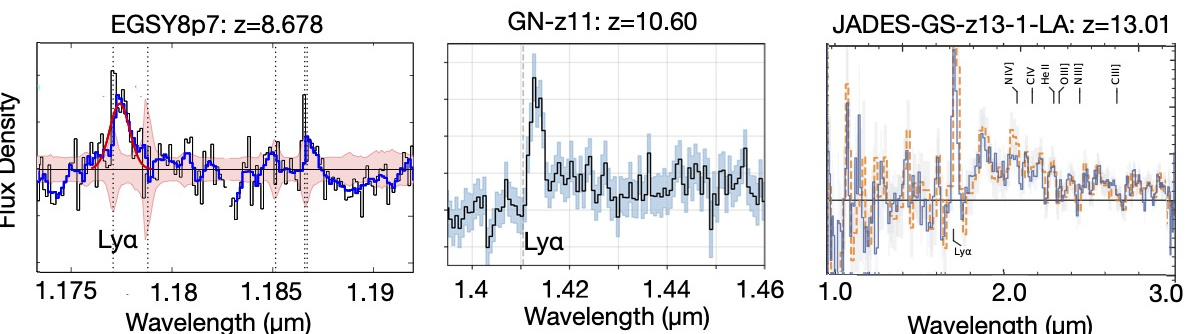,width=0.95\textwidth}}
\vspace{-0.25cm}
\caption{Spectra showing Ly$\alpha$ emission from galaxies within the
  first 600 Myr of cosmic history. The left panel presents a
  Ly$\alpha$ detection at $z=8.69$ obtained with Keck/MOSFIRE
  \citep{Zitrin2015}. The middle and right panels show {\it
    JWST}/NIRSpec detections at $z=10.6$ \citep{Bunker2023_gnz11} and
  $z=13.01$ \citep{Witstok2025Nat}. Observing Ly$\alpha$ emission this
  early is surprising given the high neutral hydrogen fraction
  inferred for the IGM, implying the presence of substantial ionized
  regions (``bubbles'') around these galaxies.\label{fig:lymanalpha}}
\end{figure*}

By measuring the Ly$\alpha$ equivalent widths (line strength relative
to the neighboring continuum) and escape fractions across many sources
\citep{Tang2023,Nakane2024,Napolitano2024,Jones2025,Saxena2023,Tang2024,Chen2026_bubbles,Napolitano2025,Kageura2025},
it is now possible to map the neutral hydrogen fraction as a function
of spatial position in the early universe and to compare it with the
distribution of ionizing sources such as galaxies or quasars.  This
comparison can now be made in many well-studied fields
\citep{Tang2024,Chen2026_bubbles,Lu2025,Nikolic2025,Martin2026}.
Figure~\ref{fig:bubbles} (from Chen et al. \citep{Chen2026_bubbles})
illustrates the three-dimensional distribution of star-forming
galaxies and highlights those systems for which the surrounding
hydrogen is sufficiently ionized for Ly$\alpha$ emission to be
visible.  {\it JWST}'s coverage of Balmer-line transitions
($n\geq3\rightarrow2$) enables direct measurements of the Ly$\alpha$
escape fractions for individual galaxies
\citep{Roy2023,Chen2024,Lin2024}.

One particularly noteworthy result in this context has been the
detection of Ly$\alpha$ emission from two galaxies at z$>$10, the
first from GN-z11 at z=10.60 \citep{Bunker2023_gnz11} and the second
from a galaxy at z$=$13.01 (Figure~\ref{fig:lymanalpha})
\citep{Witstok2025Nat}.  These detections point to the presence of
ionized bubbles as early as 300-400 Myr after the Big Bang, well
before the universe was expected to contain a significant number of
such regions.  Interestingly, even before {\it JWST}, one bright
Ly$\alpha$-emitting galaxy was identified at $z=8.69$
\citep{Zitrin2015}, pointing to the existence of ionized regions deep
in the reionization era.  These results suggest that the topology of
reionization is highly inhomogeneous, with ionized regions forming
early around the most luminous galaxies.

\section{Current Frontiers of Investigation}

Given the many transformational results from {\it JWST} to date and
the many open questions that remain, several areas are poised for
substantial progress:

\subsection{Exploration at z$\geq$15}

Reaching beyond $z=15$ would take {\it JWST} into the era it was
designed to study: the emergence of the first observable galaxies.
Even a small confirmed sample would constrain how rapidly the first
dark-matter halos accumulated gas and formed stars, becoming detectable
within 300 million years of the Big Bang.

\begin{figure*}
\centerline{
  \psfig{file=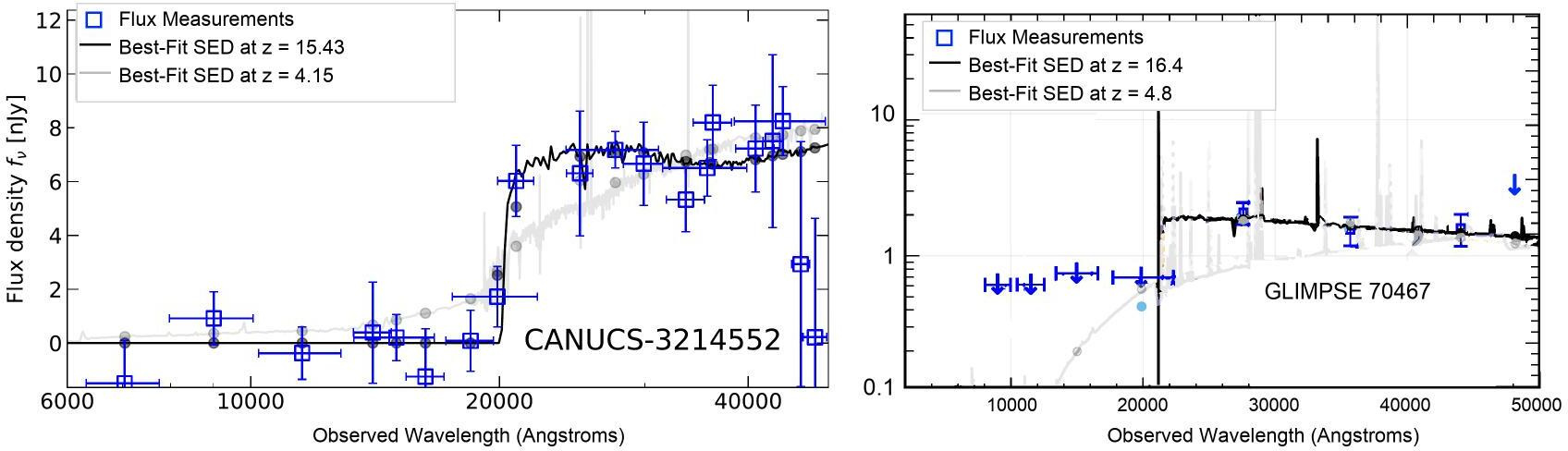,width=0.99\textwidth}}
\vspace{-0.45cm}
\caption{(\textit{Left}) Spectral energy distribution of
  CANUCS-3214552, a candidate galaxy at $z\simeq15.4$ that may lie
  beyond the current spectroscopic redshift record ($z=14.44$)
  \citep{Naidu2026}, identified in deep NIRCam imaging from the CANUCS
  + Technicolor program \citep{Asada2026}.  Blue points show the
  observed photometry; the black curve shows the best-fit template at
  $z=15.43$, compared with a much poorer low-redshift fit
  (\textit{grey curve}).  (\textit{Right}) GLIMPSE~70467, a candidate
  galaxy at photometric redshift $z=16.4\pm1.8$, found in deep NIRCam
  imaging behind the massive lensing cluster Abell S1063 ($z=0.348$)
  from the GLIMPSE program \citep{Kokorev2025_z16}, compared against a
  poorer low-redshift fit ($z=4.8\pm1.0$).  These are only two among
  many plausible $z\geq15$ candidates now being identified
  \citep{Castellano2025,PerezGonzalez2025,Gandolfi2026}. Although
  extending the high-redshift frontier will require significant {\it
    JWST} observing time, further progress is likely in the coming
  years.\label{fig:high-z-cand}}
\end{figure*}

Although early {\it JWST} observations identified many candidate
$z>15$ galaxies, including those reported by Donnan et al.
\citep{Donnan2023}, Harikane et al. \citep{Harikane2023}, and Atek et
al. \citep{Atek2023}, none has yet been spectroscopically confirmed at
$z\gtrsim15$. Such galaxies may therefore be less common than
initially inferred. This would not be surprising, since gravitational
collapse needs time to build observable galaxies.

Nevertheless, wider NIRCam surveys
\citep{Casey2023_COSMOSWeb,Williams2025} and fields magnified by
gravitational lensing \citep{Bezanson2024,Atek2025_glimpse} may allow
{\it JWST} to identify a few galaxies at $z\geq15$. Growing
medium-band coverage will assist these searches by distinguishing
genuine high-redshift galaxies from lower-redshift interlopers. Such
observations now cover the JWST Origins Field \citep{Eisenstein2025},
UNCOVER \citep{Suess2024}, the CANUCS parallel fields
\citep{Martis2025,Asada2026}, and the PRIMER and CEERS (Cosmic
Evolution Early Release Science) fields through MINERVA
\citep{Muzzin2025} and SPAM \citep{Davis2025}.

Figure~\ref{fig:high-z-cand} (\textit{left panel}) presents one
promising z$\sim$15 candidate galaxy from the CANUCS + Technicolor
program \citep{Asada2026}, while the right panel shows a candidate
z$\sim$16 galaxy found in sensitive GLIMPSE data lensed by the
foreground cluster Abell S1063 \citep{Kokorev2025_z16}.  Multiple
ongoing NIRSpec programs aim to enlarge the samples at $z>10$ and
$z>14$ \citep{Casey2025_jwst,Harikane2025_jwst}.

\subsection{High Spatial-Resolution Study of Faint z$>$6 Galaxies}

\begin{figure*}
\centerline{
  \psfig{file=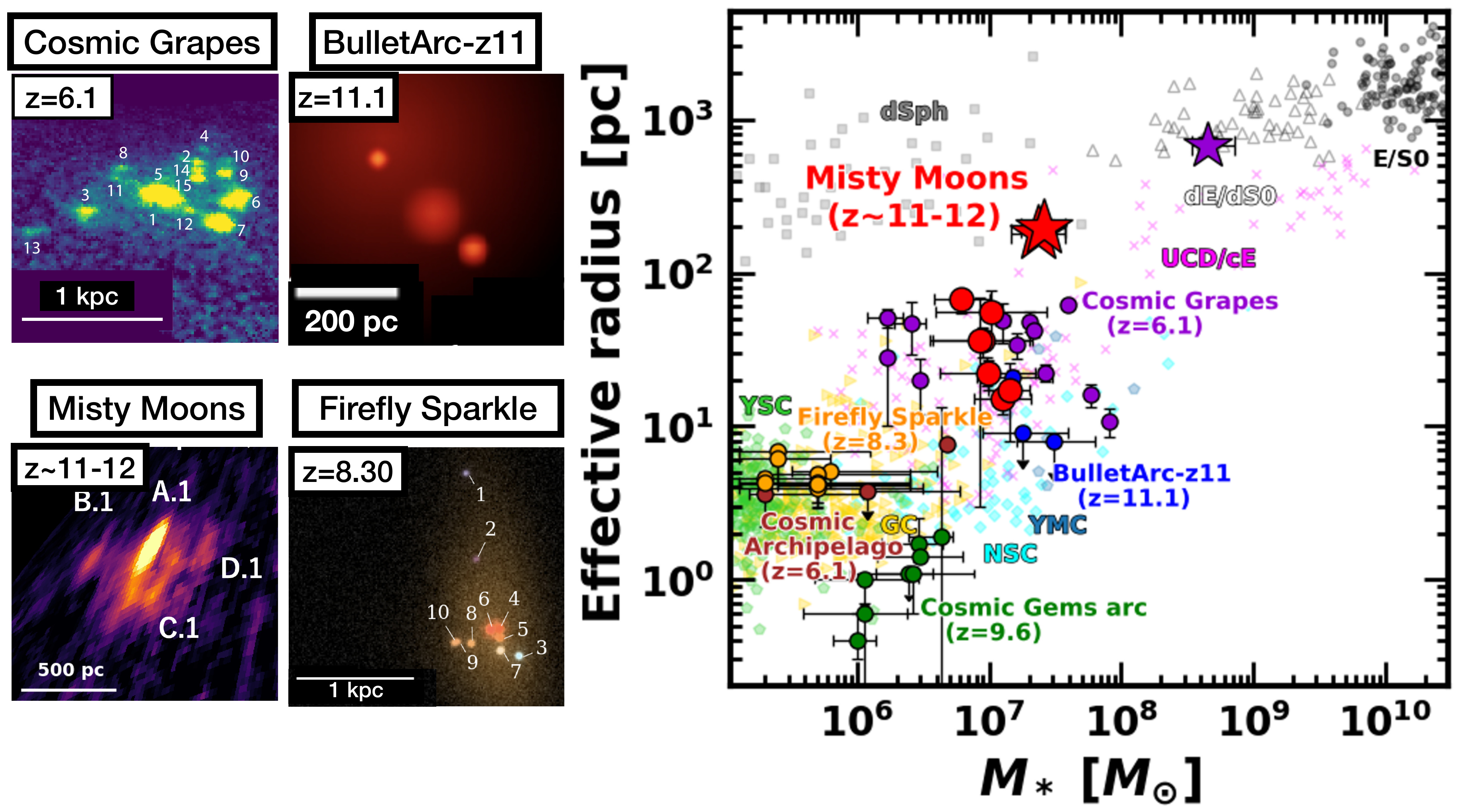,width=0.9\textwidth}}
\vspace{-0.35cm}
\caption{(\textit{Left}) High spatial-resolution images of four
  gravitationally-lensed star-forming systems within the first Gyr of
  cosmic history -- Cosmic Grapes ($z=6.1$), BulletArc-z11
  ($z=11.1$), Misty Moons ($z\sim11$--12), and Firefly Sparkle
  ($z=8.30$) \citep{Mowla2024,Fujimoto2025_cosmicgrapes,Bradac2025,Nakane2025}.
  These reconstructions account for the lensing models of the
  foreground clusters. Lensing boosts the effective {\it JWST}
  resolution by $\sim50\times$, enabling studies of structures on
  scales of $<1$ kpc and in some cases $<200$ pc. The images suggest
  that star formation in early galaxies occurs in clustered complexes
  rather than smooth disk-like structures. (\textit{Right}) Effective
  radius versus stellar mass for the individual clumps identified in
  these systems (colored points), compared with local star clusters
  and galaxies: young star clusters (YSC), young massive clusters
  (YMC), nuclear star clusters (NSC), globular clusters (GC),
  ultra-compact dwarfs and compact ellipticals (UCD/cE), dwarf
  spheroidals (dSph), dwarf elliptical/S0 galaxies (dE/dS0), and
  elliptical/S0 galaxies (E/S0). The clumps occupy a region of
  parameter space intermediate between star clusters and dwarf
  galaxies. This panel is from Nakane et al. \citep{Nakane2025}.
  \label{fig:lensing}}
\end{figure*}

Another exciting area of exploration has been the study of normal
star-forming galaxies at high magnification using gravitational
lensing by foreground clusters.  Several such programs have now
resolved highly magnified galaxies across $z\simeq6$--12
\citep{Fujimoto2025_cosmicgrapes,Mowla2024,Messa2025,Hsiao2024,Bradac2025,Nakane2025}
(Figure~\ref{fig:lensing}). At high magnifications $\mu$ $>$50, it is
possible to detect both continuum and line emission from galaxies
forming stars at rates as low as $0.02\,M_{\odot}\,\mathrm{yr}^{-1}$
\citep{Messa2025}.

In addition to boosting the apparent brightness of sources,
gravitational lensing enables observations at spatial resolutions more
than $\sim 50\times$ higher than would otherwise be possible.  This
allows structures to be probed down to spatial scales of $<5$ pc,
revealing that star-forming galaxies at high redshift are not smooth
systems, but instead consist of extremely compact ($\sim$1–100 pc),
spatially separated, star-forming clumps
\citep{Fujimoto2025_cosmicgrapes,Mowla2024,Messa2025,Hsiao2024,Bradac2025,Nakane2025}.
This provides us with direct insight into the formation and evolution
of star clusters in the early universe
\citep{Adamo2024,Bradley2025,Vanzella2026}.

\subsection{Characterization of z$>$10 Galaxies using NIRSpec, MIRI, and ALMA --
  Metallicities, ISM Conditions, and Dynamical Properties}

\begin{figure*}
\centerline{
  \psfig{file=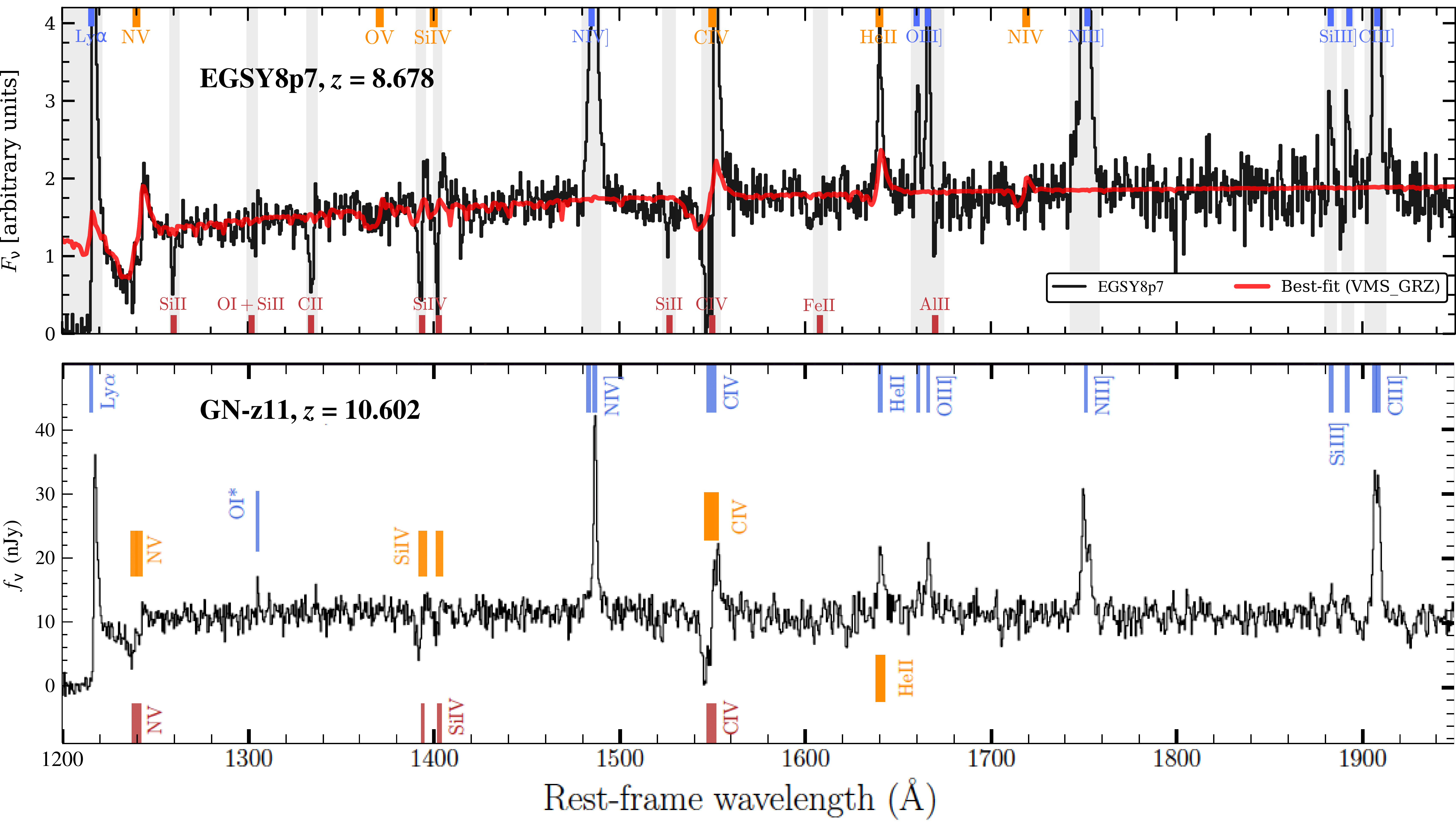,width=0.99\textwidth}}
\vspace{-0.3cm}
\caption{Very sensitive, high-spectral-resolution {\it JWST}/NIRSpec
  spectroscopy of the rest-UV light from EGSY8p7 (CEERS-1019) and
  GN-z11, two especially bright star-forming galaxies at $z=8.678$ and
  $z=10.602$, respectively. These spectra provide exceptionally
  high-S/N information on nebular emission (\textit{blue}), stellar
  winds (\textit{yellow}), and interstellar absorption
  (\textit{red}). The stellar-wind features include P-Cygni profiles,
  the combined emission-and-absorption signature of gas flowing
  outward from massive stars. The spectra therefore provide detailed
  probes of the most massive stars in these galaxies and illustrate
  the potential of sensitive {\it JWST} spectroscopy for
  characterizing the earliest galaxies. Adapted from figures in
  Marques-Chaves et al. \citep{MarquesChaves2026_spurs} and Chen et
  al. \citep{Chen2026b}.\label{fig:spurs}}
\end{figure*}

Deep NIRSpec, MIRI $\lambda>5\,\mu\textrm{m}$, and far-IR ALMA
observations are driving key advances for z$>$10 galaxies.  Deep
($>$30 hour), high spectral resolution ($R=\lambda/d\lambda$$>$1000)
NIRSpec spectroscopy is now underway for z$>$9 galaxies across key
extragalactic legacy fields, including the 200-hour SPURS
(SPectroscopic Ultra-deep Reionization-era Survey) large program
\citep{Mason2025_spurs,Chen2026a,Chen2026b} and even deeper programs
\citep{Stark2025_jwst,Senchyna2025_jwst}, extending results from the
$>$800-hour JADES spectroscopic campaign
\citep{CurtisLake2026,Scholtz2026}. These exposures directly
probe ISM conditions (density and abundances), stellar populations,
gas outflows, and the dominant sources of ionizing photons (massive
stars vs AGN) in early galaxies (Figure~\ref{fig:spurs}).

MIRI spectroscopy beyond $z=10$ is essential because it reaches
familiar optical signatures of ionized oxygen and hydrogen shifted
beyond NIRSpec's range \citep{Kendrew2015}. These features probe how
episodically stars form and how chemically enriched the gas has become
\citep{Kennicutt1998,Lee2009,Calzetti2013}. Current observations cover
only a small sample of exceptionally bright confirmed galaxies at
$z\simeq10$--14
\citep{Hsiao2024_MIRI,AlvarezMarquez2026,AlvarezMarquez2025,Zavala2025,Helton2026}. Most
show clear oxygen and hydrogen emission, implying heavy-element
abundances of roughly one tenth solar, with no clear evidence for
extreme ongoing bursts. Observations of $\gtrsim10$ additional $z>10$
galaxies in the coming year
\citep{AlvarezMarquez2025_jwst,Endsley2025_jwst,Schouws2025_jwst}
should significantly sharpen constraints on early galaxy assembly.

ALMA is likewise advancing our understanding of luminous $z>10$
systems. While {\it JWST} probes stellar and rest-UV emission, ALMA
traces dust buildup and obscured star formation. The bright [O
  III]$_{88}$ line, combined with ALMA's sensitivity and velocity
resolution, provides powerful constraints on galaxy kinematics
\citep{Schouws2025,Scholtz2025}. A high-S/N ($7\sigma$) detection in
JADES-GS-z14-0 at $z=14.1793$ was achieved in just 3 hours
\citep{Schouws2025,Carniani2025}.  Comparisons of [O III]$_{88}$ with
      [O III]$_{5007}$ from MIRI also constrain electron density,
      implying $>500\,\mathrm{cm}^{-3}$ for JADES-GS-z14-0
      \citep{Schouws2025,Carniani2025,Helton2026}. Beyond this source,
            [O III]$_{88}$ has been detected in seven other galaxies
            at $z>8$, spanning $z=8.31$--12.33
            \citep{Tamura2019,Hashimoto2018,Heintz2023,Algera2026_PHOENIXpilot,Witstok2026,Zavala2024}. Approved
            observations of 15 more luminous systems are underway
            through the PHOENIX ALMA Large Program, marking a
            transition from isolated detections to samples capable of
            revealing population trends.

\subsection{Topology of Reionization}

Another rapidly advancing area is reionization topology: the
three-dimensional distribution, sizes, and connectivity of ionized
regions within otherwise neutral hydrogen.  As discussed in
\S\ref{sec:reionization}, Ly$\alpha$ is an especially sensitive probe
because neutral hydrogen readily scatters these photons, whereas an
ionized bubble allows them to reach us.  Most reionization models
predict that reionization broadly proceeds ``inside-out'' from the
highest density regions of the universe to lower density regions
\citep{Barkana2001,Furlanetto2004,Mellema2006,Iliev2006,Choudhury2009,Hutter2018,Trebitsch2021,Gnedin2022,Kannan2022},
but in detail reality may be more complicated because those regions of
the universe with the most sources of ionizing photons also contain
the densest gas which most readily recombines.

Existing studies of Ly$\alpha$ emission versus environment seem
consistent with the general expectation that reionization begins from
the densest regions and proceeds to lower density regions
\citep{Chen2024,Chen2026_bubbles} (Figure~\ref{fig:bubbles}). Drawing
firm conclusions has been challenging, however: protoclusters like the
$z=7.9$ one behind Abell 2744 show evidence for significant neutral
gas \citep{Chen2024,Witten2025,Fudamoto2025}, while many galaxies in
relatively low-density regions appear to sit in ionized gas
\citep{Chen2024,Chen2026_bubbles}.  With increasing spectroscopic
redshift \citep{Meyer2025} and Ly$\alpha$ coverage
\citep{Lin2025_jwst} of well-studied extragalactic fields, as well as
spectroscopic coverage of more z$>$6 protoclusters in the rest-$UV$,
the situation should rapidly improve.

\section{Summary}

Understanding how galaxies formed during the first billion years of
cosmic history has long been a central goal of observational
cosmology. The {\it James Webb Space Telescope} has reshaped this
field by enabling the detection and characterization of galaxies at
previously inaccessible redshifts, probing epochs when the universe
was only a few hundred million years old.

This review places {\it JWST} discoveries within the broader context
of early-universe studies. The {\it Hubble} and {\it Spitzer} Space
Telescopes identified galaxies out to z$\sim$11, but were limited by
wavelength coverage and spectroscopic sensitivity. {\it JWST} extends
observations into the near- and mid-infrared with greater collecting
area and spatial resolution, enabling robust identification via
Ly$\alpha$ breaks and direct spectroscopic confirmation.

These advances are driven by {\it JWST}’s instrument suite: deep
NIRCam/NIRISS imaging and slitless spectroscopy for selection, NIRSpec
multi-object spectroscopy for redshifts and physical characterization,
and MIRI for older stars, AGN, and spectroscopy beyond
5$\mu$m. Together, these capabilities enable both the discovery and
detailed characterization of large high-redshift galaxy samples.

Early surveys have expanded the census of galaxies at $z>10$,
revealing populations more numerous and luminous than many models
predicted.  While these results suggest rapid early growth and
efficient star formation, uncertainties remain due to challenges in
interpreting the bright sources themselves.  Spectroscopic follow-up
has been essential for confirming these sources and ascertaining their
properties.

Spectral energy distributions constructed from {\it JWST} imaging data
provide us a view of the stars, dust, and AGN in these galaxies, while
spectroscopy constrains ionization conditions, gas reservoirs,
chemical enrichment, and feedback. Current {\it JWST} observations
point to galaxies with high gas fractions, intense radiation fields,
and highly dynamic interstellar media, often dominated by massive
stars and strong nebular emission. High spatial resolution imaging, in
some cases enhanced by gravitational lensing, is probing the internal
structure of these galaxies on $<20$ pc scales, revealing compact
morphologies and clumpy star formation during early assembly.

Despite much progress, important open questions remain regarding the
efficiency, intermittency, and initial mass function of star formation
at the highest redshifts, interstellar-medium conditions, masses, and
structure of early galaxies,
and the topology of reionization, particularly given the unexpectedly
large number of luminous galaxies. Resolving these open questions will
require deeper spectroscopy including to $>$5$\mu$m, larger
statistical samples, lensing magnification, and coordinated
multiwavelength observations combining {\it JWST} with facilities such
as ALMA.

\section*{Acknowledgements}

Rychard Bouwens gratefully acknowledges the many helpful suggestions
and feedback from Pascal Oesch, Rohan Naidu, Jorryt Matthee, Hiddo
Algera, and Renske Smit.  Rychard is appreciative to Fengwu Sun and
collaborators for sharing some of their WFSS redshift results in
advance of publication and also to Kasper Heintz and Joris Witstok for
including him in the April 2026 “Charting Cosmic Dawn” meeting in
Copenhagen, which provided a valuable overview of recent developments
in the field.  Rychard is also grateful to Richard Ellis for inviting
this review and for many useful informal suggestions, and to the
referees for their constructive comments, which significantly improved
the manuscript.

\section*{Notes on contributor}

Rychard Bouwens is an astrophysicist at Leiden Observatory
specializing in the formation and evolution of galaxies in the early
universe. Rychard is a leading expert in extragalactic cosmology and
early galaxy formation, whose extensive work leverages data from the
{\it Hubble}, {\it Spitzer}, and {\it James Webb Space
  Telescopes}. His research focuses on understanding how stars, gas,
and dust rapidly formed in the earliest massive galaxies, addressing
some of the most fundamental questions about cosmic dawn. Rychard has
led and contributed to major international programs, including large
{\it HST} and ALMA programs targeting high-redshift galaxies, and has
consistently been recognized among the world’s most highly cited
researchers.

\IfFileExists{tfnlm.bst}
  {\bibliographystyle{tfnlm}}
  {\bibliographystyle{unsrtnat}}
\bibliography{CP}

\end{document}